\documentclass[11pt]{article}
\pdfoutput=1

\usepackage{jheppub}
\usepackage{amsmath}
\usepackage{amssymb}
\usepackage{dcolumn}
\usepackage{bm}
\usepackage{multirow}
\usepackage{blkarray}
\usepackage{feynmp}        
\usepackage{slashed}                             
\usepackage{hyperref,graphicx}           
\usepackage{url}
\usepackage{caption}
\usepackage{subcaption}
\usepackage{multirow}
\usepackage{units}
\usepackage{xcolor}
\usepackage{graphicx}
\usepackage{tikz}

\def\c{\checkmark}

\begin{document}

\title{Anatomy of the $tthh$ Physics at HL-LHC}
\author{Lingfeng Li$^a$, Ying-Ying Li$^b$, Tao Liu$^b$}
\affiliation{
$^a$Jockey Club Institute for Advanced Study, The Hong Kong
University of Science and Technology, Hong Kong S.A.R.\\
$^b$Department of Physics, The Hong Kong
University of Science and Technology, Hong Kong S.A.R.}

\abstract{
The $tthh$ production at colliders contain rich information on the nature of Higgs boson. In this article, we systematically studied its physics at High-Luminosity Large Hadron Collider (HL-LHC), using exclusive channels with multiple ($\geq 5$) $b$-jets and one lepton ($5b1\ell$), multiple ($\geq 5$) $b$-jets and opposite-sign di-lepton ($5b2\ell$), same-sign di-lepton (SS2$\ell$), multiple leptons (multi-$\ell$), and di-tau resonance ($\tau\tau$). The scenarios analyzed include: (1)~the $tthh$ production in Standard Model; (2)~the $tthh$ production mediated by anomalous cubic Higgs self-coupling and $tthh$ contact interaction; (3)~heavy Higgs ($H$) production with $tt H \to tthh$; and (4)~pair production of fermionic top partners ($T$)  with $T T \to tthh$. To address the complication of event topologies and the mess of combinatorial backgrounds, a tool of Boosted-Decision-Tree was applied in the analyses. The $5b1\ell$ and SS2$\ell$ analyses define the two most promising channels, resulting in slightly different sensitivities. For the non-resonant $tthh$ production, a combination of these exclusive analyses allows for its measurment in the SM with a statistical significance $\sim 0.9\sigma$ (with $S/B > 1 \%$), and may assist partially breaking the sensitivity degeneracy w.r.t. the cubic Higgs self-coupling, a difficulty usually thought to exist in gluon fusion di-Higgs analysis at HL-LHC. These sensitivities were also projected to future hadron colliders at 27 TeV and 100 TeV. For the resonant $tthh$ productions, the heavy Higgs boson in type II Two-Higgs-Doublet-Model could be efficiently searched for between the mass thresholds $2 m_h < m_H < 2 m_t$ and even beyond that, for relatively small $\tan\beta$ (vacuum alignment), while the fermionic top partners in composite Higgs models could be probed for up to $\sim 1.5$ TeV and $\sim 1.7$ TeV, for Br$(T\to th)=25\%$ and $50\%$, respectively.      
}

\maketitle

\section{Introduction}
\label{sec:intro}

The $tthh$ production at colliders contain rich information on the nature of Higgs boson. Together with gluon fusion ($gg\to hh$), vector-boson-fusion (VBF) and vector boson association (VBA), the $tthh$ production defines the set of main mechanisms for di-Higgs production in Standard Model (SM) (for discussions on their next-leading-order (NLO) cross sections at hadron colliders, see, e.g.,~\cite{Frederix:2014hta}). These di-Higgs productions can be applied to detect Higgs self-couplings and hence  probe for the nature of electroweak phase transition (EWPT)~\cite{Noble:2007kk}. Their measurements therefore have been established as one of the main tasks at High-Luminosity Large Hadron Collider (HL-LHC) ($\mathcal{L}=3$\,ab$^{-1}$@14 TeV)~\cite{Atlas:2019qfx}, future Higgs factory (e.g., International Linear Collider (ILC)~\cite{Djouadi:2007ik}), and future hadron colliders such as High-Energy LHC (HE-LHC) ($\mathcal{L}=15$\,ab$^{-1}$@27 TeV)~\cite{Zimmermann:2018wdi}, Super-proton-proton Collider(SppC) ($\mathcal{L}=30$\,ab$^{-1}$@100 TeV)~\cite{CEPC-SPPCStudyGroup:2015csa} and Future Circular Collider (FCC)-$hh$  ($\mathcal{L}=30$\,ab$^{-1}$@100 TeV)~\cite{Golling:2016gvc,Contino:2016spe}.  

Due to its relatively large cross section at hadron colliders, the $gg\to hh$ production has received the most attentions in literatures so far. The analyses performed by the ATLAS and CMS collaborations~\cite{ATL-PHYS-PUB-2018-053,CMS:2018ccd} show that the cubic Higgs self-coupling could be measured at HL-LHC, with a precision of $\sim \mathcal O(1)$ at $2\sigma$ C.L.. Recently, the analysis of the $gg\to hh \to bb \gamma\gamma$ channel was extended to 27 TeV, indicating that the cubic Higgs self-coupling could be measured with a precision $< \mathcal O(1)$~\cite{Goncalves:2018yva, Bizon:2018syu,Homiller:2018dgu,ATL-PHYS-PUB-2018-053}. Yet, because of the strategic difference in addressing the backgrounds, pileups, etc., the sensitivities obtained in literatures vary by a factor of two or three~\cite{Cepeda:2019klc}. At 100 TeV, the $gg\to hh \to bb \gamma\gamma$ analysis has been extensively carried out (see, e.g.,~\cite{Yao:2013ika,Barr:2014sga,Azatov:2015oxa,He:2015spf,Contino:2016spe,Goncalves:2018yva,Chang:2018uwu,Bizon:2018syu}). It shows that a precision of $10\%$ for the cubic Higgs self-coupling measurement is possible, and could be further improved, using the detector tailered for a 100 TeV machine~\cite{Contino:2016spe,Goncalves:2018yva}.

\begin{figure}
\centering
\includegraphics[scale=0.7]{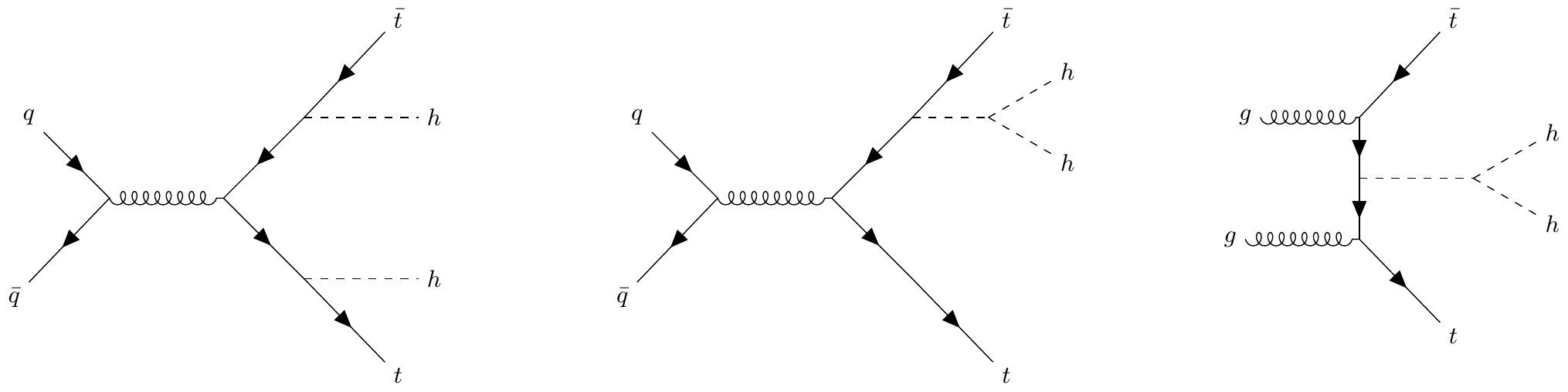}
\caption{Tree-level Feynman diagrams for the $tthh$(SM) production. }
\label{fig:fey}
\end{figure}

Despite of this, the $tthh$ production may play a special role in measuring the Higgs self-couplings. As is well-known, the gluon fusion di-Higgs production involves destructive interference between Feynman diagrams with and without the cubic Higgs self-coupling. This may result in a sensitivity degeneracy at HL-LHC between two disjointed regions of the cubic Higgs self-coupling: one near the SM and another one far away~\cite{ATL-PHYS-PUB-2018-053}, though this degeneracy could be broken with kinematics at future hadron colliders~\cite{Cepeda:2019klc}.  Differently, the relevant interference in the $tthh$ production is constructive. Thus there might be a chance to break this degeneracy at HL-LHC either completely or partially with the assistance of this channel.  Second, because of its higher energy threshold of production, the $tthh$ cross section increases faster than the $gg \to hh$ one, as the beam energy $\sqrt{s}$ increases. As a matter of fact, their difference in the SM is reduced from a factor $\sim 35$ at 14 TeV to $\sim 14$ at 100 TeV~\cite{Frederix:2014hta}). This indicates that the $tthh$ measurements may gain more in sensitivity at future hadron colliders, compared to $gg\to hh$. More than that, the di-Higgs productions can be applied to probe for the quartic Higgs self-coupling~\cite{Maltoni:2018ttu,Liu:2018peg}, since it may renormalize the cubic Higgs self-coupling and contribute to the form factor of the relevant vertices at NLO. Yet, as is discussed in~\cite{Liu:2018peg}, a combined analysis of multiple di-Higgs productions is valuable in reducing the dependence of the generated limits on renormalization scheme, a non-physical effect resulting from incomplete treatment of higher-order corrections. The $tthh$ measurements may serve for such a purpose.    

Aside from measuring the Higgs self-couplings, the $tthh$ channel can also serve as a probe for theories of electroweak symmetry breaking (EWSB) such as Composite Higgs Models(CHM) and supersymmetry. In the CHM, there typically exists an extended top sector which couples with the Higgs boson. With the heavy degrees of freedom being integrated out at a cutoff of strong dynamics, a contact interaction of $tthh$ can be generated~\cite{Agashe:2004rs,Dib:2005re,Contino:2006qr,Grober:2010yv,Giudice:2007fh}. This operator will contribute to the non-resonant $tthh$ production at tree level and hence mediate its signal rate at colliders (for studies on the impact of the $tthh$ contact interaction for the $gg\to hh$ kinematics, see, e.g.,~\cite{Chen:2014xra}). The CHM may contribute to the resonant $tthh$ production as well. For example,  the top partner ($T$) can be pair produced in the CHM with both of them decaying as $T \to th$. Actually, this is one of the major channels for its searches at LHC~\cite{Aaboud:2018xuw}. Besides that, the resonant $tthh$ production may arise from Two Higgs Doublet Model (THDM) of, e.g., type II, including Minimal Supersymmetry Standard Model (MSSM). In this scenario, a heavy CP-even Higgs resonance $H$ can be produced in association with a pair of top quarks, and subsequently decays into a pair of Higgs bosons. This channel may become important for the $H$ searches, if $2 m_h < m_H < 2 m_t$, with a relatively small $\tan\beta$ ($i.e.$, vacuum alignment). Indeed,  in this nearly decoupling limit the on-shell decay of $H \to tt$ is turned off, while the couplings of $H$ with gauge bosons tend to be suppressed. It has been shown numerically that the decay of $H\to hh$ could be dominant in this parameter region~\cite{Djouadi:2015jea}. A branching ratio of Br$(H\to hh) \sim \mathcal O(10\%)$ can be extended to $m_H \sim 400$ GeV and $\tan\beta \sim 5$~\cite{Djouadi:2015jea}. Note, the $tthh$ contact interaction can be also generated in these models by integrating out heavy bosonic degrees of freedom.  

However, the $tthh$ sensitivities at colliders have been analyzed in very few papers so far. The first analyses at HL-LHC were pursued in~\cite{Englert:2014uqa,Liu:2014rva}, using a specific channel $tthh \to tt bbbb$ and a cut-based method. Subsequently, the ATLAS experimentalists re-analyzed its sensitivity, in a similar setup but with more aggressive kinematic cuts~\cite{ATL-PHYS-PUB-2016-023}. It showed that a statistical significance of $S/\sqrt{B}\sim 0.35$ could be achieved, by combining two significances in quadrature: the bin with 5 $b$-tags ($0.23\sigma$, $S/B\sim 0.3\%$) and the bin with $\geq 6$ $b$-tags ($0.26\sigma$, $S/B\sim 1.1\%$). Yet, such analyses could be further optimized. The $tthh$ production is characterized by a complicated topology. The jet multiplicity in signal and, strongly correlated with this, the combinatorial backgrounds make the event reconstruction very challenging. It has been known for a while, such a difficulty could be addressed to some extent by a tool of Boosted-Decision-Tree (BDT). Actually, the BDT method has been extensively applied in the measurements of top physics at LHC~\cite{top2018}, such as the ones of $tt+V$ \cite{Aad:2015eua}, $tt+h\to bb$ \cite{Aaboud:2017rss}, and $tt+h \to \gamma\gamma/ZZ^*$ \cite{Aaboud:2018urx}. It has also been suggested to use for the searches of heavy Higgs bosons which are produced in association with top or bottom quarks at hadron colliders~\cite{Hajer:2015gka,Craig:2016ygr}. More than that, almost all studies so far were focused on the specific channel of $tthh \to tt bbbb$, given its relatively large signal rate. The channels with, e.g., same-sign di-lepton and multiple leptons, were largely ignored, which could be valuable also, because of their relatively clean backgrounds. 

Motivated by these considerations, in this article we will systematically analyze the sensitivities of measuring the $tthh$ physics at HL-LHC, using the tool of BDT, in the following physics scenarios: 
\begin{enumerate}

\item (non-resonant) the $tthh$ production in the SM (denoted as $tthh{\rm (SM)}$) (their tree-level Feynman diagrams are shown in Figure~\ref{fig:fey}).

\item (non-resonant) the $tthh$ production mediated by an anomalous cubic Higgs self-coupling and the $tthh$ contact interaction; 

\item (resonant) heavy Higgs production with $tt H \to tthh$; 

\item (resonant) pair production of fermionic top partners with $T T \to tthh$. 

\end{enumerate}
Limited by statistics, we will focus on five exclusive channels with multiple ($\geq 5$) $b$-jets and one lepton, multiple ($\geq 5$) $b$-jets and opposite-sign di-lepton, same-sign di-lepton, multiple leptons, and di-tau resonance. The $tthh$ measurements are expected to gain more at future hadron colliders, given the increased $tthh$ cross section, improved kinematics such boost, and potentially larger luminosity. We will leave their study in a separate paper~\cite{LLLL}. 

We organize this article as follows. We will introduce analysis strategy in Section \ref{sec:analysis}. The sensitivities of measuring the non-resonant and resonant $tthh$ physics at HL-LHC are presented respectively in Section \ref{sec:result} and Section \ref{sec:resultresonant}. To qualitatively measure the potential of future hadron colliders, we also project the sensitivities of probing for the anomalous cubic Higgs self-coupling and the $tthh$ contact interaction to 27 TeV (HE-LHC) and 100 TeV (SppC and FCC-hh) in Section \ref{sec:result}. We summarize our studies and discuss potential directions to explore in Section \ref{sec:conclusion}. Technical details on our analyses are provided in Appendices.

\section{Analysis Strategy}
\label{sec:analysis}

\subsection{Generation of Signal and Background Events}
\label{sec:simulation}

In this study, we use MadGraph 5~\cite{Alwall:2011uj} and Pythia 8~\cite{Sjostrand:2007gs} to generate the signal and background events at leading order. The heavy components ($t,~W,~Z$ and $h$) in each event decay inclusively in either MadGraph 5~\cite{Alwall:2011uj}, Pythia 8~\cite{Sjostrand:2007gs} or MadSpin~\cite{Artoisenet:2012st}.

The detector simulation is performed using Delphes 3~\cite{deFavereau:2013fsa}, with a detector performance suggested for the HL-LHC runs. All leptons are required to have $p_T>\unit[10]{GeV}$ and be isolated. The isolation is defined with a cone $\Delta R =0.2$, and the net $p_T$ within it ($i.e.$, the total $p_T$ excluding the contribution from target lepton $l$) being smaller than 0.2$p_T(\ell)$. 
Jets are clustered using anti-$k_t$ algorithm with $\Delta R=0.4$. All of them are required to have $p_T> \unit[20]{GeV}$. 

The tagging of $b$-jets plays an important role in this study. We define its efficiencies with the working points of  MV2 $b$-tagging algorithm~\cite{ATL-PHYS-PUB-2016-012}. Explicitly, we choose the tight, moderate and loose $b$-tagging efficiencies to be 60\%, 77\% and 85\%, respectively. We also analyze the sensitivities with a softened tight $b$-tagging efficiency of 70\%, to see the potential impact of the $b$-tagging efficiency for the $tthh$ sensitivities at HL-LHC. Unless specified, the results based on the tight and softened tight $b$-tagging efficiencies will be presented. $\tau$ jets are tagged with an efficiency $\sim 60\%$, with a mis-tagging rate $\sim 1.5\%$ for jets with $p_T >20$~GeV and $|\eta|<2.3$. This setup is consistent with the HL-LHC expectation~\cite{ATL-PHYS-PUB-2015-045}. 

For event reconstruction in BDT, we reconstruct fat jets using anti-$k_t$ algorithm in a second clustering exercise, with $\Delta R=1.0$ and $p_T > 200$ GeV. The fat jets are pruned using a jet trimming algorithm~\cite{Krohn:2009th}, with $\Delta R$=0.3 and $p_T$ fraction of trimming being 0.05. This to some extent can weaken the impact of the pileups~\cite{ATLAS:2012kla} which are not turned on in this simulation. The MET reconstruction can be also smeared by high pileups in detector ~\cite{JETM-2016-012}. Our analyses do not significantly rely on the MET measurements. But still we mimic this impact by smearing the MET with a vector of $\unit[50]{GeV}$. 

\begin{table}[th]
\centering
\begin{tabular}{|c|c|c|}
\hline 
Channels & $\sigma$(14 TeV)[fb] & Generated Events  \\ 
\hline
\multicolumn{3}{|c|}{Non-resonant $tthh$ signal} \\
\hline
$tthh{\rm (SM)}$ & 0.981 (NLO)  & $2\times10^6$ \\ 
\hline
$y^4$, $y^2\kappa^2$, $c_{t}^2$, $y^3\kappa$, $y^2c_{t}$, $y\kappa c_{t}$ & / & ${5.5\times10^5}$ for each \\
\hline
\multicolumn{3}{|c|}{Resonant $tthh$ signal} \\
\hline
$ttH \to tthh$, $m_H$=300~GeV  & / & ${1.05\times10^6}$ \\ 
$ttH \to tthh$, $m_H$=500~GeV  & /   & ${5.5\times10^5}$ \\ 
\hline
$TT\to tthh$, $m_T$=1.5~TeV & 2.4$\times$ BR$(T\to th)^2$  & ${1.05\times10^6}$\\
$TT\to tthh$, $m_T$=1.75~TeV & 0.65 $\times$ BR$(T\to th)^2$  & ${5.5\times10^5}$\\
$TT\to tthh$, $m_T$=2~TeV &  0.19$\times$ BR$(T\to th)^2$  & ${1.05\times10^6}$\\ 
\hline
\multicolumn{3}{|c|}{Background of multi-tops} \\
\hline 
$4t$ & 11.8  & ${5.5\times10^5}$ \\ 
\hline 
\multicolumn{3}{|c|}{Backgrounds of di-tops} \\
\hline
$tt4b$ & 370  &${1.05\times10^6}$\\ 
\hline 
$tt2b2c$  & 103  &${1.05\times 10^6}$ \\ 
\hline 
$ttVbb$& 27.6  &${6\times10^5}$   \\ 
\hline 
$tthbb$ & 15.6  & ${5.5\times10^5}$\\ 
\hline 
$ttVV$ & 14.6  &${3\times10^5}$\\ 
\hline
$tthZ$ &$1.55$ &  ${3\times10^5}$\\ 
\hline 
\end{tabular} 
\caption{ Event generation for signals and their main backgrounds. The di-boson background $VV$ includes $WW$, $WZ$ and $ZZ$. LO cross sections (except the $tthh$(SM) cross section) are listed.
}
\label{tab:bkglist}
\end{table}

For the non-resonant $tthh$ analysis, the signal events are generated based on a simplified Lagrangian~\cite{Contino:2012xk}
\begin{equation}
\mathcal{L} \supset -y \frac{m_t}{v} t\bar{t}h -\kappa \frac{1}{3!}\frac{3 m_h^2}{v}h^3 -  c_t \frac{1}{2!} \frac{m_t}{v^2} t\bar{t}hh ~.
\end{equation}
This Lagrangian is reduced to the SM if the dimensionless top Yukawa coupling $y$, the scaled cubic Higgs self-coupling $\kappa$ and the $tthh$ Wilson coefficient $c_t$ are valued as
\begin{eqnarray}
y\equiv \frac{y_{tth}}{y_{tth}^{\rm SM}}=1, \ \  \kappa\equiv \frac{ \lambda_{hhh}}{\lambda_{hhh}^{\rm SM}}=1,  \ \  c_{t} = 0 \ .
\end{eqnarray}
Given that the top Yukawa coupling can be measured separately (e.g., using the $tth$ production) at HL-LHC~\cite{Cepeda:2019klc} with a precision better than the other two, we fix its value to be one in this article. Totally seven signal samples are generated. The first one is defined by the $tthh$(SM) production. The BDT performance will be illustrated mainly using this sample. This sample will be also used as one of the backgrounds for the resonant $tthh$ analyses. Another six samples are denoted by a set of coupling products $\{y^4, y^2\kappa^2, c_{t}^2, y^3\kappa, y^2c_{t}, y\kappa c_{t}\}$. They represent contributions of the six terms in the LO $tthh$ cross section. These samples will be used to analyze the sensitivities, given the $\{\kappa, c_t\}$ values, by reweighting their relative contributions. For the resonant $tthh$ analysis, totally five samples are generated: two for the scenario of heavy Higgs boson, and three for the scenario of fermionic top partner. The samples in each scenario are characterized by different resonance masses. 

The major backgrounds shared by the non-resonant and the resonant $tthh$ productions include four-top ($4t$) and di-top + $X$, with $X= 4b$, $bbcc$, etc. We will not include the contributions of di-top + $X$ with $X= 4c, \ 2b/2c + 2j, \ 4j$. By approximately matching the simulations with fiducial cross section of the $tt$ with multiple $b$-jets measured at ATLAS~\cite{Aaboud:2018eki}, we find that in the one lepton + light jets channel ($\geq  5b$, $\geq  2j$), the $tt+4b/2b2c$ sample explains $\sim 70\%$ of the expected event number from QCD. Here the MV2 $b$-tagging algorithms~\cite{ATL-PHYS-PUB-2016-012} with an efficiency of $60\%$ are applied. So, we will define a $K$ factor with $K = \frac{\sigma_{fid}}{\sigma_{sim}}=1.44$ for the $tt+4b/2b2c$ backgrounds to represent this discrepancy\footnote{Besides the NLO effect of $tt+4b/2b2c$, the $tt + 4c, \ 2b/2c2j, \ 4j$ faked events may also contribute to this discrepancy. We will simply absorb them into the overall $K$ factor.}. As for the other backgrounds, their cross sections will be universally scaled by an NLO $K$ factor 1.2. This is comparable, e.g., for the $4t$ background, to the numbers suggested in literatures (see, e.g.,~\cite{Sirunyan:2017roi}).  The details on the event generation of signals and backgrounds is summarized in Table~\ref{tab:bkglist}.

\subsection{Selection Rules}
\label{ssec:selectionrule}

In this study, the $tthh$ events are exclusively selected for five analyses after a preselection:  $\geq 1$ lepton with $p_T > 15$ GeV and $|\eta|<2.5$; $\geq 2$ tight $b$-jets (70\% tagging efficiency) and $\geq 1$ moderate $b$-jets with $p_T>20$ GeV and $|\eta|<2.5$. Then each of them is analyzed using BDT. The five exclusive analyses include:

\begin{itemize}
\item Multiple $b$-jets ($ \geq  5$) + one lepton ($5b1\ell$). $\geq 5$ tight $b$-jets are required. This analysis targets on the decay mode of $hh \to 4b$, and was suggested to be one of the most promising channels for the $tthh$ measurement~\cite{Liu:2014rva,Englert:2014uqa}.

\item Multiple $b$-jets ($ \geq  5$) + opposite sign di-lepton ($5b2\ell$). Similar to the previous one, $\geq 5$ tight $b$-jets are required. This analysis also targets on the decay mode of $hh \to 4b$.

\item Same-sign di-lepton (SS$2\ell$). Of the two same-sign leptons, the leading one needs to have $p_T>17$ GeV and the subleading one $p_T > 14$ GeV. The events with a third lepton will be vetoed. Additionally, $\geq 4$ $b$-jets are required, among which at least three are tight, or two tight and one moderate. This analysis targets on the decay mode of $h\to VV^*$. This is reminiscent of the SS$2\ell$ search of the top quark associated single top production at LHC~\cite{Aaboud:2017jvq,Aaboud:2018xpj} which plays an important role in the measurement of top Yukawa coupling.

\item Multiple leptons ($\geq  3$) (multi-$\ell$). The same requirements for $b$-jets are applied, as the ones for the SS$2\ell$ analysis. $\geq 3$ leptons with $p_T > 10$ GeV are required. This analysis also targets on the decay mode of $h\to VV^*$.

\item Di-tau jets ($\tau\tau$). The same requirements for $b$-jets are applied, as the ones for the SS$2\ell$ analysis. Both $\tau$ jets need to have $p_T>20$ GeV.  This analysis targets on the decay model of $h \to \tau\tau$.

\end{itemize}

Being lack of precise knowledge on the trigger setup at HL-LHC, we use the ATLAS 2018 HLT triggers~\cite{trigger} as a reference in the analyses. Below are the relevant ones:
\begin{enumerate}
\item Single lepton trigger: one isolated lepton with $p_T>25$ GeV.
\item Two lepton trigger: two isolated leptons. Electrons need to have $p_T>17$ GeV and muons $p_T>14$ GeV.
\item Three muon trigger: at least three isolated muons with $p_T> \unit[6]{GeV}$.
\item Four $b$-jet trigger: at least four $b$-jets (77\% tagging efficiency), or two $b$-jets (85\%) + two $b$-jets (70\%), with each of them having $p_T>35$ GeV.
\item $3b+j$ trigger: at least three $b$-jets (70\%) and one extra jet, with each having $p_T>35$ GeV.
\item Di-$\tau$ trigger: two $\tau$-jets with the leading(subleading) one having $p_T>35(25)$ GeV.
\item $\tau+\ell$ trigger: one $\tau$-jet with $p_T>25$ GeV and one $e(\mu)$ with $p_T>17(14)$ GeV.  
\end{enumerate}
We will demonstrate the potential sensitivities combining the samples defined by these triggers. The information of the sample selection for each exclusive analysis is summarized in Table~\ref{tab:trigger}.
\begin{table}[th]
\centering
\begin{tabular}{|c|c|c|c|c|c|}
\hline 
Analysis & 5$b$1$\ell$ & 5$b$2$\ell$ & SS2$\ell$ & Multi-$\ell$ & $\tau\tau$\\
\hline
Triggers & $\cup \{1, 4, 5\}$ & $\cup \{1, 2, 4, 5\}$ & $\cup \{2\}$ & $\cup \{1, 2, 3, 4, 5\}$& $\cup \{1, 4, 5, 6, 7\}$\\ 
\hline
\end{tabular}
\caption{Sample selection in the five exclusive analyses.}
 \label{tab:trigger}
\end{table}

\subsection{Boosted Decision Tree }
\label{ssec:BDT}
\begin{figure}[ht]
\centering
\includegraphics[scale=1.3]{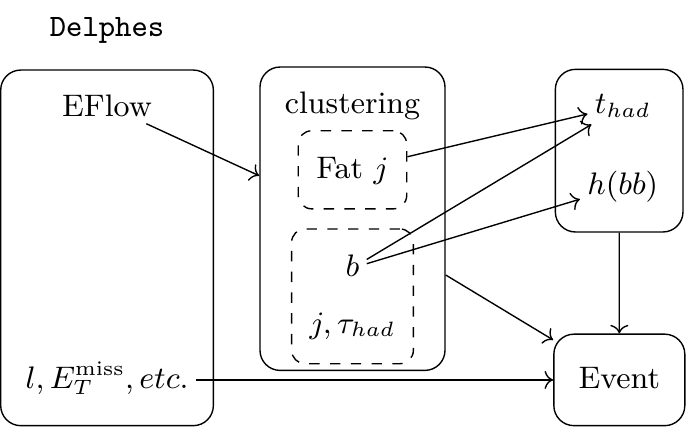}
\caption{Workflow of the BDT-based event reconstruction.}
\label{fig:workflow}
\end{figure}

To improve the sensitivity, we apply a BDT tool in the analyses, based on the TMVA package\cite{Hocker:2007ht}. The workflow of the BDT-based event reconstruction is presented in Figure~\ref{fig:workflow}. For event reconstruction in our analyses, two intermediate-object taggers, $i.e.$, one hadronic top tagger and one Higgs tagger, are constructed from lower-level inputs.

The top tagger mainly targets on boosted or relatively boosted top quarks in the $tthh$ events. It is constructed using the information of paired fat jet and $b$-jet, with $\Delta R< 2$ between them. Here the fat jets and the $b$-jets are generated from two separate clustering exercises. Besides $p_T$, mass and $\eta$ of the fat jet and the $b$-jet, subjetness ($\tau_{1-4}$)~\cite{Plehn:2009rk} and track number of the fat jet, tagging level (tight/moderate/loose) of the $b$-jet, and their separation in azimuthal angle $\Delta \phi$ are also used as the variable inputs of the tagger. For each fat jet, only the pairing with a $b$-jet which scores highest by the BDT is kept for subsequent data processing. The four momentum of the reconstructed top is defined as that of the fat jet, if the fat jet has $\Delta R < 1$ w.r.t. the $b$-jet or its mass is larger than $70$ GeV (typical upper limit for the mass of the trimmed $W$-jets), and as the total four momentum of the paired fat jet and $b$-jet for the other cases. The tagger is trained using single top events ($t+1j$) with $p_T > 150$ GeV for the top quark, against the SM backgrounds.

The Higgs tagger is constructed using the information of paired $b$-jets with $160 {\rm \ GeV}>m_{bb}>90$ GeV. The variable inputs include $p_T$, mass and $\eta$ of each paired $b$-jet, their tagging levels (tight/moderate/loose) and their separation in azimuthal angle $\Delta \phi$. The four momentum of the reconstructed Higgs is defined as the total four momentum of the paired $b$-jets. The tagger is trained using the $hZ \to bb\ell\ell$ events with $p_{T}>25$~GeV for the leptons and $p_{T}>20$~GeV for the $b$-jets.

In each event at most two BDT top quarks and three BDT Higgs bosons will be kept which are sorted by their BDT scores. These top and Higgs candidates, together with the other objects in the events, are used for the event reconstruction. The variables include not only the ones which have been widely accepted ~\cite{Aad:2015eua,Aaboud:2017rss,Aaboud:2018urx}, but also the ones encoding the kinematics of the reconstructed objects and their correlation with each other (e.g., their separations $\Delta R_{h_i,h_j}$,  $\Delta R_{h_i,t_j}$, etc., and their invariant mass $m_{h_i,h_j}$, $m_{h_i,t_j}$, etc.) and with the other objects in the event. To characterize their overlapping, we flip the sign of the correlation variables if the two reconstructed objects share some common elements (e.g., if a tagged top $t_i$ and a tagged Higgs $h_j$ share the same $b$-jet).  A list of these variables are summarized in Table~\ref{tab:BDTvariables}. Note, in the non-resonant $tthh$ analyses we take the variables of separation for event reconstruction. They are replaced with the variables of invariant mass in the resonant analysis.

\section{Non-Resonant $tthh$ Analysis}
\label{sec:result}

\subsection{The $tthh$(SM) Production}

\begin{table}[th]
\resizebox{\textwidth}{!}{
\begin{tabular}{|c|c|c|c|c|c|c|c|}
\hline 
 & No cut & Preselection & 5$b$1$\ell$ & 5$b$2$\ell$ & SS2$\ell$ & Multi-$\ell$ & $\tau\tau$\\
\hline\hline
$tthh$ & 2.9e3 & 7.37e2 & 50.9 (97.2)  & 6.1 (12.0) & 14.6 (15.7) & 8.6 (9.2) & 3.6 (3.8)\\
\hline
$tt4b$ & 1.1e6 & 1.79e5 & 6.56e3 (1.31e4) & 664 (1.30e3) & 212 (223) & 115 (121) & 94.1 (95.1)\\
$tt2b2c$ & 3.1e5 & 4.28e4 & 621 (1.73e3) & 59.4 (163) & 38.0 (42.4) & 24.1(26.8) & 43.6 (48.6)\\
$ttVV$ & 4.4e4 & 3.64e3 & 20.7 (52.7) & 3.5 (6.4) & 51.8(60.9) & 32.4 (36.5)& 3.1 (3.9) \\
$4t$ & 3.54e4 & 1.30e4 & 350 (804) & 68.3 (152) & 592 (635) & 307 (324) & 59.8 (64.2) \\
$ttbbV$ & 8.29e4 & 1.54e4 & 353 (765) & 47.8 (105) & 114 (124) & 203 (221) & 22.2 (24.2) \\
 $ttbbh$ &4.68e4 & 1.04e4 & 608 (1.15e3) & 69.0 (136) & 91.0 (98.0) & 53.4 (56.2) & 24.2 (25.9)\\
 $tthZ$ & 4.65e3 & 881 & 28.1 (58.5) & 4.1 (9.1)  & 8.8 (9.5) & 18.5 (19.9) & 2.3 (2.5) \\
 \hline
Total & 1.6e6 & 2.65e5 & 8.53e3 (1.76e4) & 918 (1.88e3) & 1.11e3 (1.19e3) & 753 (806) & 249 (265) \\
 \hline
 \multicolumn{3}{|c|}{$\sigma_{\rm cut}$} & 0.46 (0.62) & 0.17 (0.23) & 0.39 (0.40) & 0.28 (0.29)& 0.20 (0.20)\\
 \hline
 \multicolumn{3}{|c|}{$(S/B)_{\rm cut} (\%)$ } & 0.42 (0.40)  &  0.47 (0.55)  & 1.1 (1.1)  & 0.9 (0.9) & 1.1 (1.1) \\
 \hline
\multicolumn{3}{|c|}{ $\sigma_{\rm BDT}$} & 0.59 (0.79) & 0.21 (0.30) & 0.45 (0.46) &  0.33  (0.35) & 0.21 (0.21) \\
 \hline
 \multicolumn{3}{|c|}{$(S/B)_{\rm BDT} (\%)$} & 1.2 (1.0) & 1.3 (1.6) &  1.6 (1.6) & 1.6 (1.9) & 1.6 (1.6) \\
 \hline
\multicolumn{3}{|c|}{ $\sigma_{\rm com}$} & \multicolumn{5}{c|}{0.86 (1.04)} \\
\hline
\end{tabular}
} 
\caption{Cut flow of the $tthh$(SM) signal and its major backgrounds at HL-LHC. In the five exclusive analyses, the numbers outside (inside) the brackets are based on a tight (softened tight) $b$-tagging efficiency of 60\% (70\%). All significances are calculated against the background-only hypothesis.}
 \label{tab:cutflow1}
\end{table}
To illustrate its effectiveness, we apply the strategy developed in Section~\ref{sec:analysis} to analyze the $tthh$(SM) first, starting with a tight $b$-tagging efficiency of $60\%$. The cut flow of the signal and backgrounds before the BDT analysis is summarized in Table~\ref{tab:cutflow1}. We take into account the $K$ factor in calculating statistical significance and the $S/B$ value. For the $5b 1\ell$ and $5b2\ell$ analyses, the backgrounds are dominated by $tt4b$. Their cut-based significances at HL-LHC are $\sim 0.5\sigma$ and $\sim 0.2\sigma$, respectively, with both $S/B$ values being of sub-percent level. For the SS$2\ell$ and multi-$\ell$ analyses, the main backgrounds are $4t$, $tt4b$ and $ttbbV/h$. Their cut-based significances at HL-LHC are $\sim 0.4\sigma$ and $\sim 0.3\sigma$, respectively. The SS$2\ell$  sensitivity performance is relatively better than the multi-$\ell$ one. Different from the $gg\to hh\to 4W$ channel~\cite{Ren:2017jbg}, where the SS$2\ell$ analysis suffers  
a lot from the di-boson backgrounds ($i.e.$, $VV$, $hV$, etc.), yielding a sensitivity relatively lower than the multi-$\ell$ one, the relevant di-boson backgrounds for the SS$2\ell$ analysis of $tthh$ ($i.e.$, $ttVV$, $tthZ$, etc.) can be efficiently suppressed with a cut of multiple $b$-jets. As for the $\tau\tau$ analysis, its main backgrounds are relatively diverse, yielding a cut-based significance comparable to that of $5b2\ell$. The $S/B$ values of these three analyses are approximately $\sim 1\%$.

\begin{figure}[th]
\begin{center}
\includegraphics[scale=0.35]{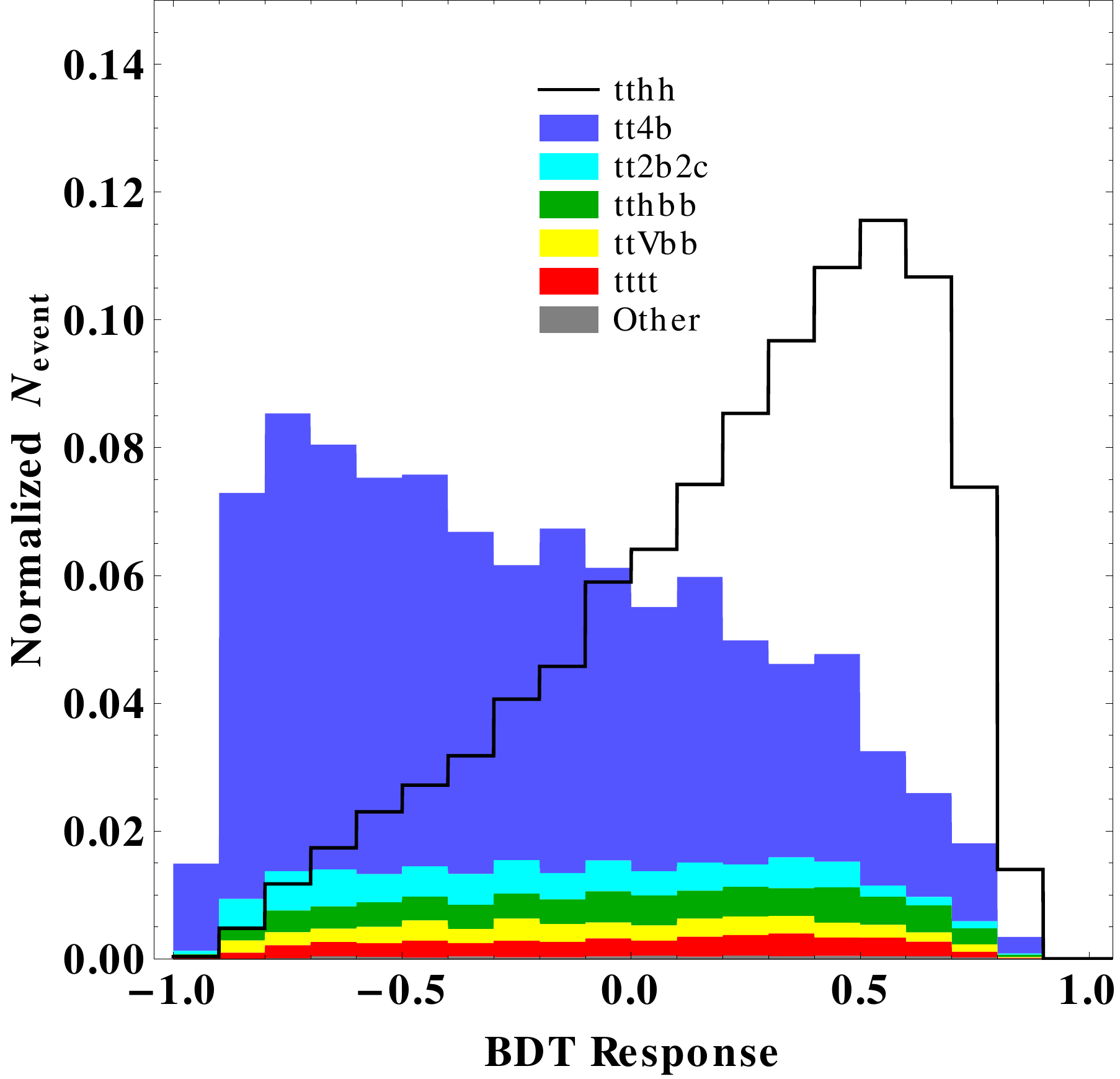} \ \ \ \ \ \
\includegraphics[scale=0.35]{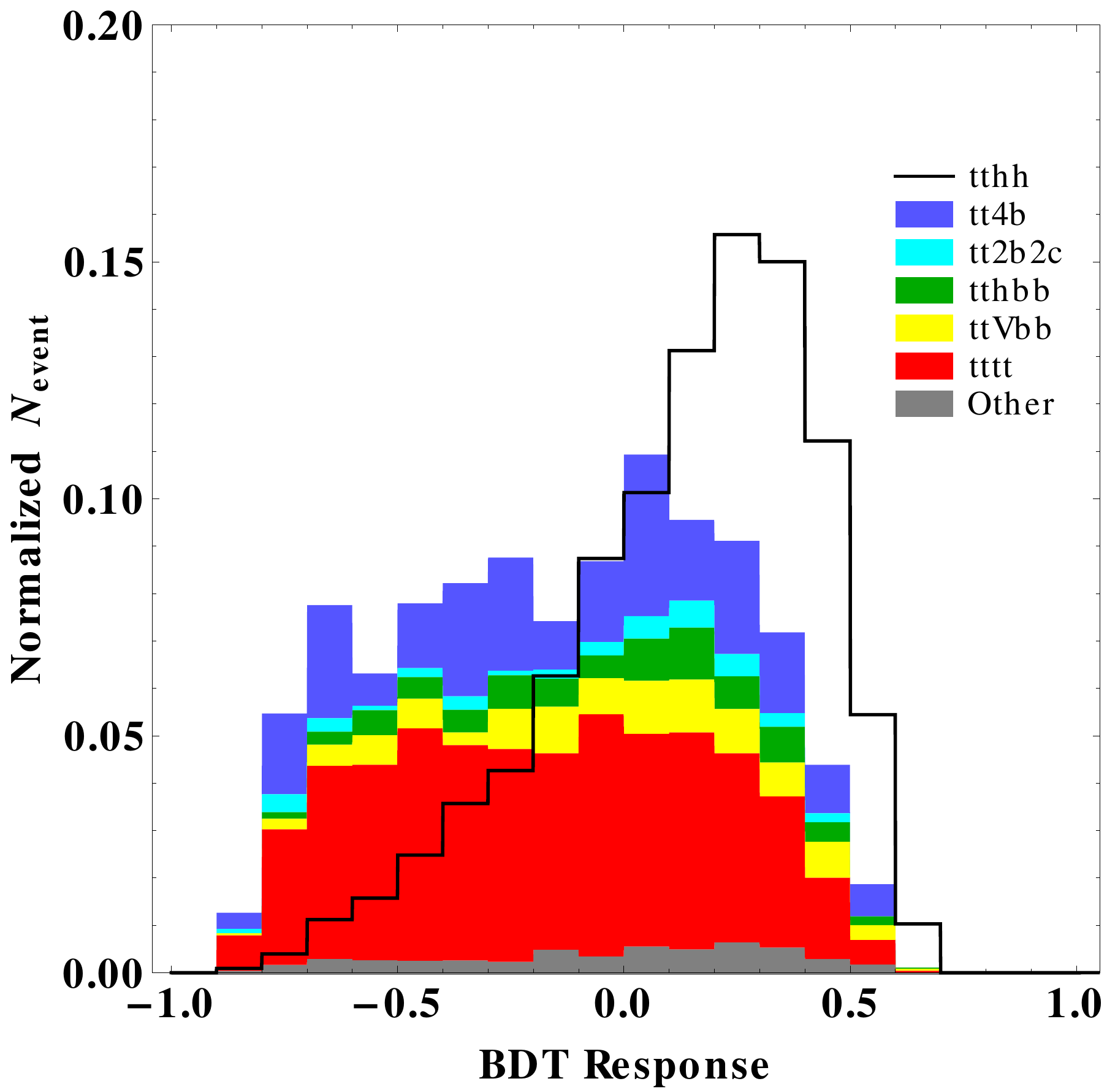} 
\end{center}
\caption{Normalized event number versus BDT response, with a tight $b$-tagging efficiency of 60\%. Left: the $5b1 \ell$ analysis. Right: the SS$2\ell$ analysis. }
\label{fig:ChannelHistograms}
\end{figure}

\begin{figure}[th]
\begin{center}
\includegraphics[scale=0.31]{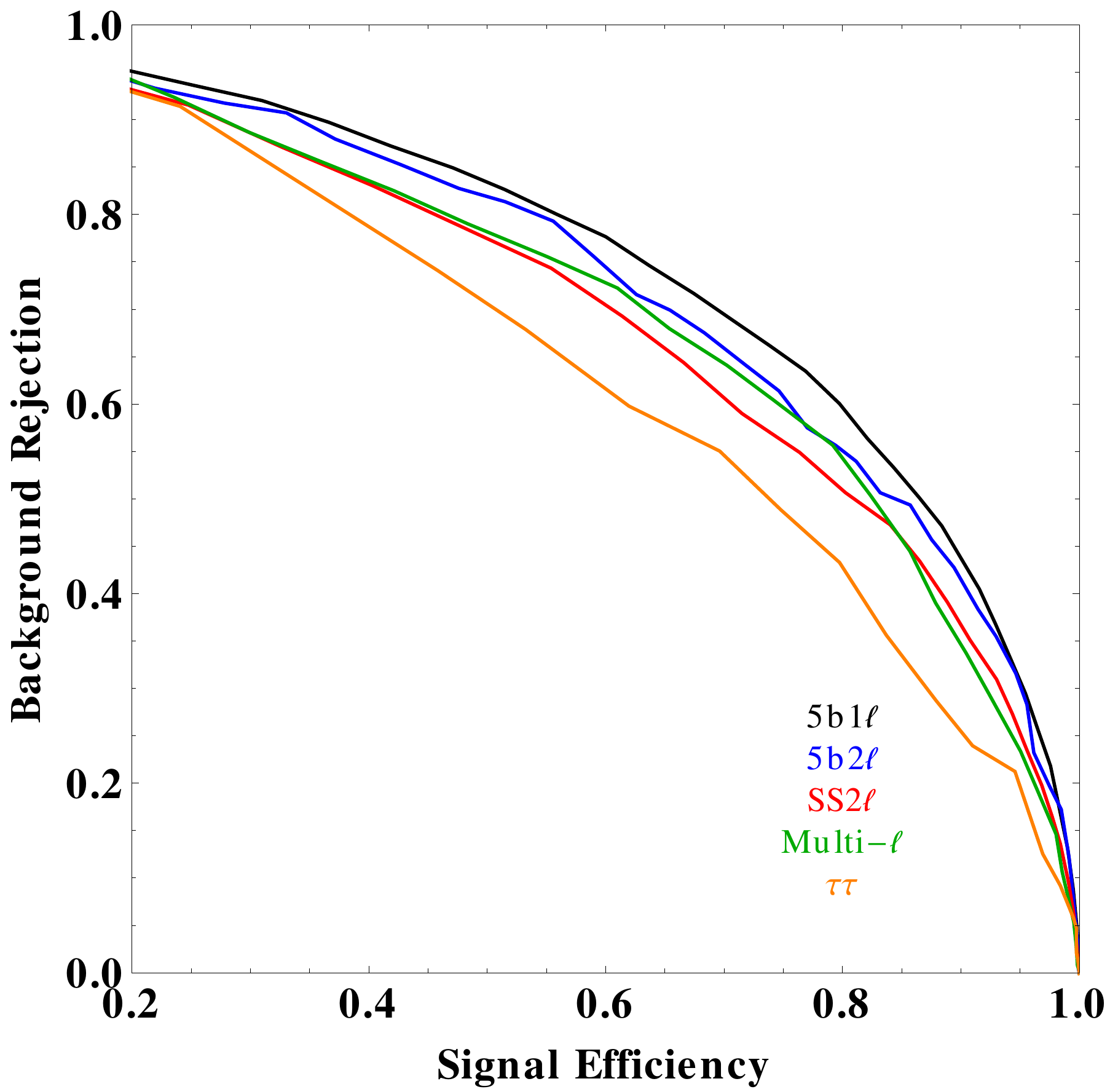} \ \ \ \ \ \ \ \
\includegraphics[scale=0.31]{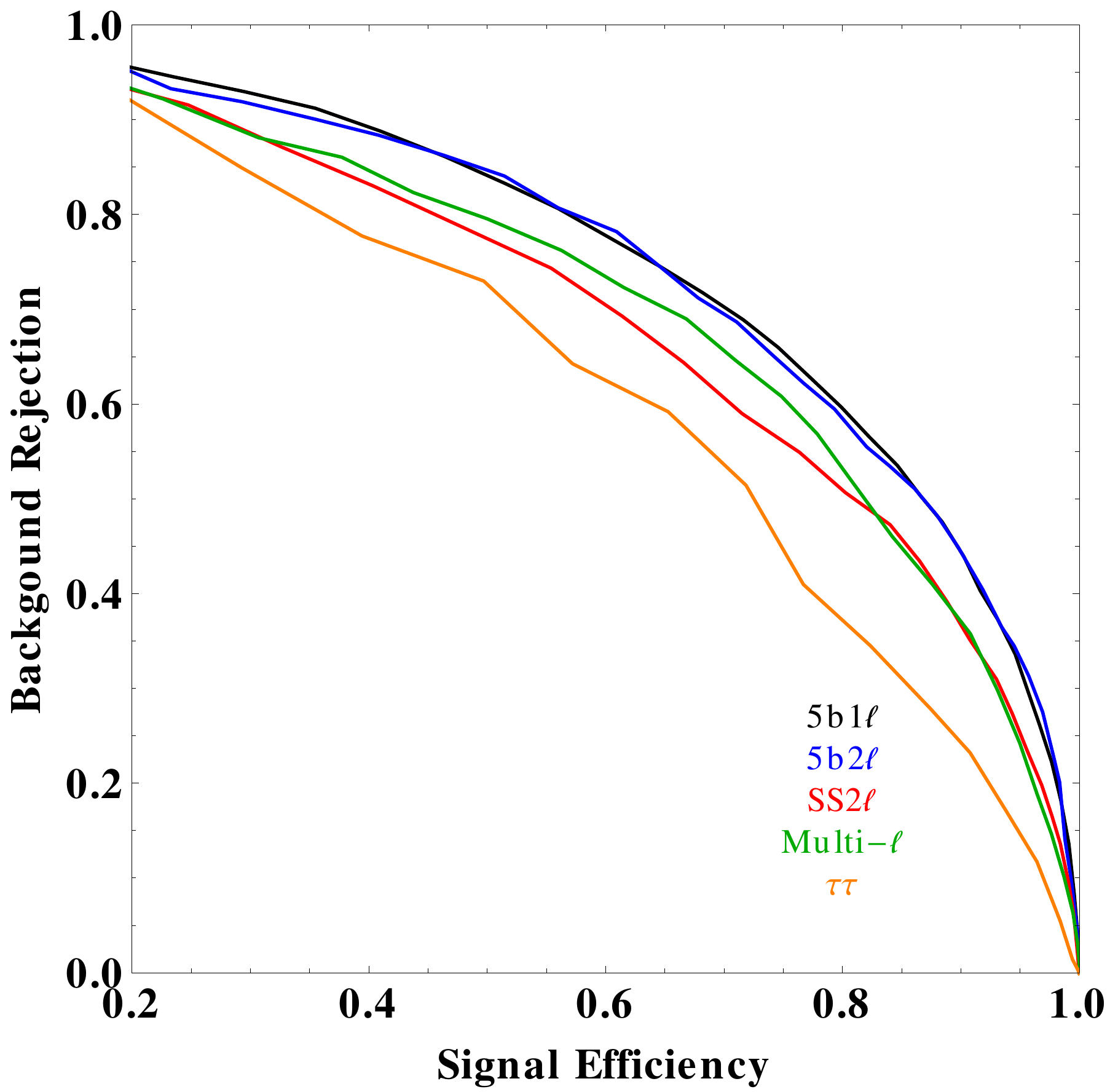}  \\
\bigskip
\includegraphics[scale=0.31]{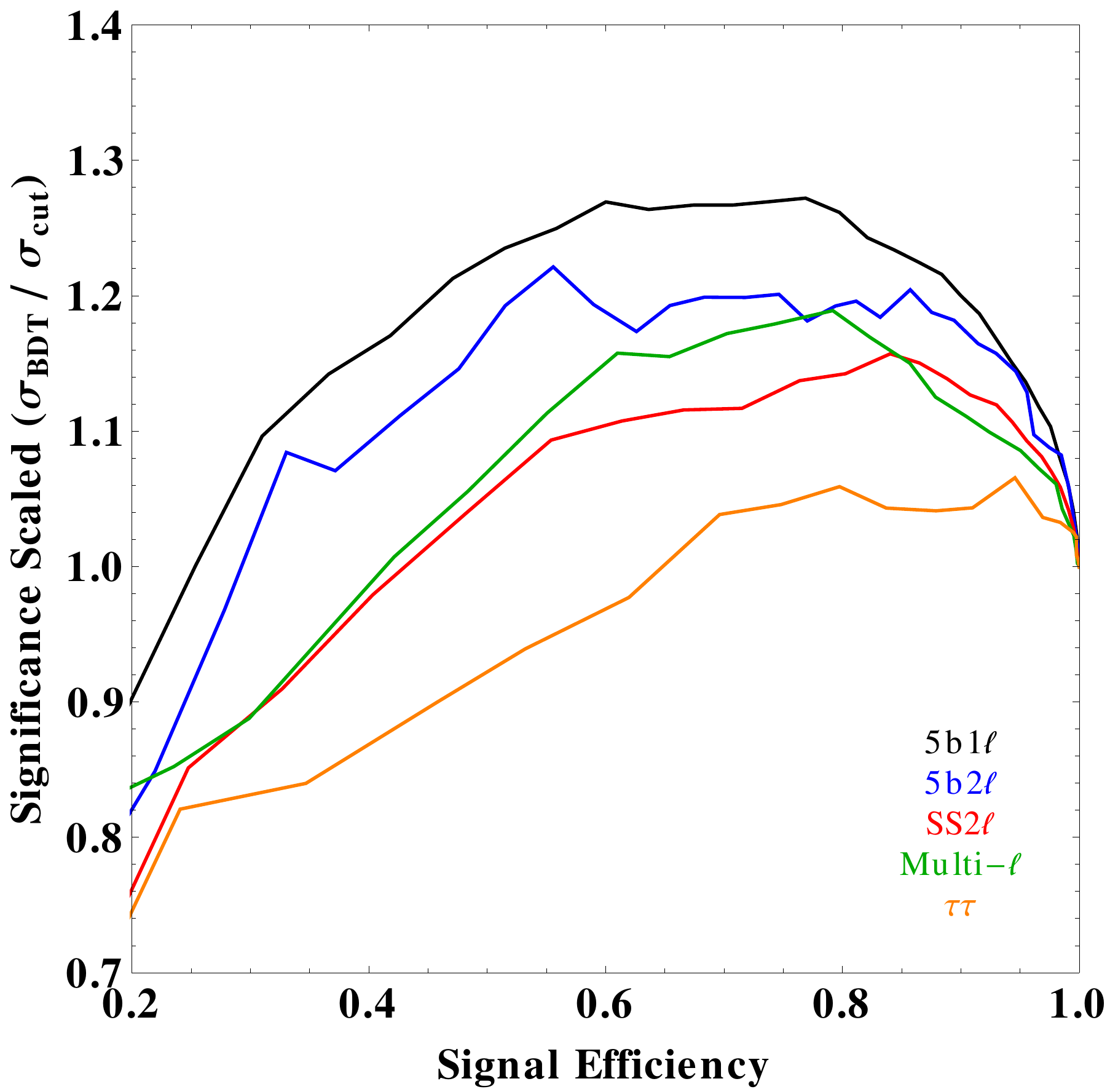} \ \ \ \ \ \ \ \
\includegraphics[scale=0.31]{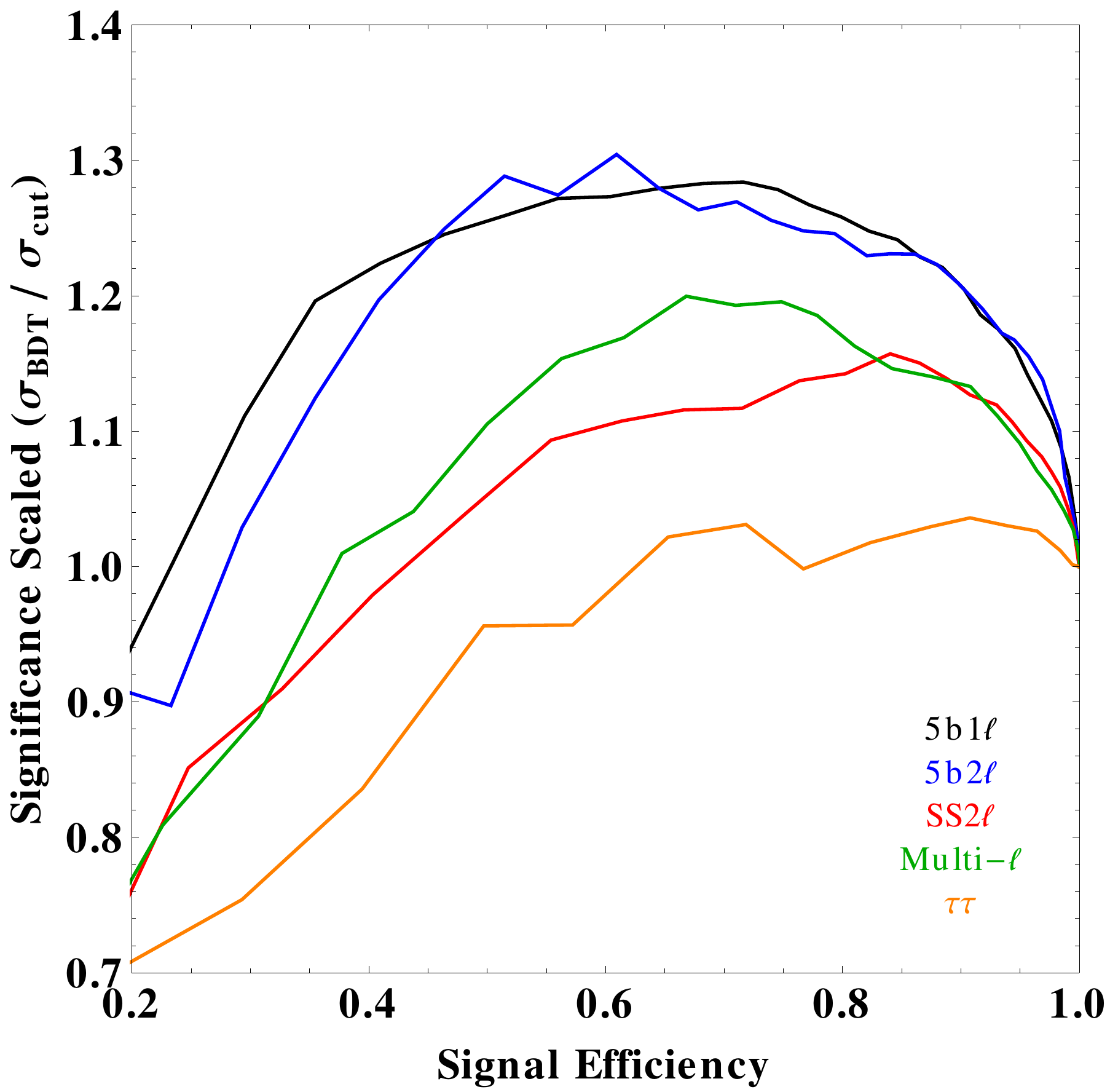}\\
\bigskip
\includegraphics[scale=0.31]{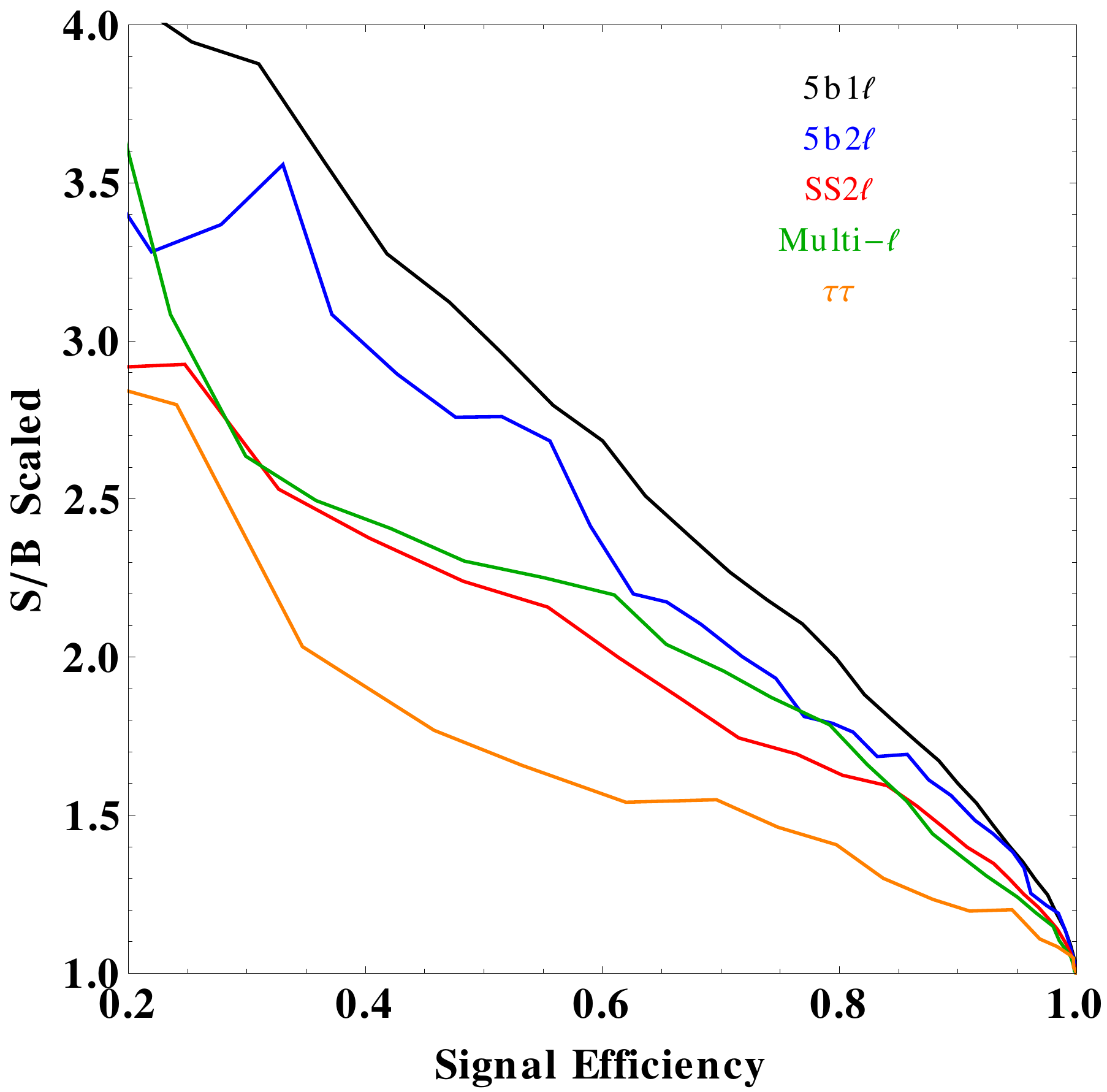} \ \ \ \ \ \ \ \
\includegraphics[scale=0.31]{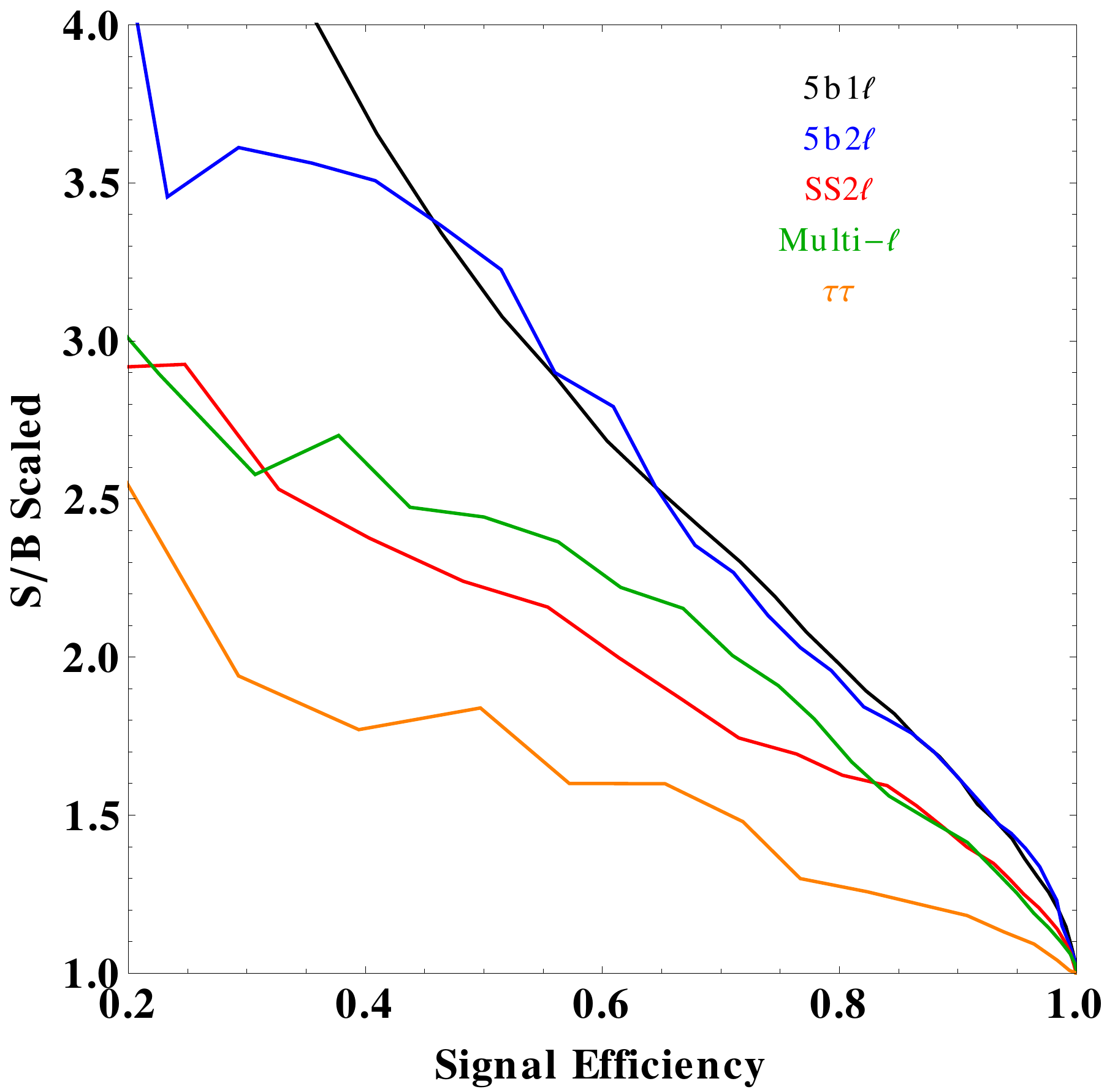}
\end{center}
\caption{ROCs (top), gain of statistical significance (middle) and gain of $S/B$ (bottom) as a function of signal efficiency. The latter two are calculated by scaling the BDT results with their respective cut-based constant values. Left: tight $b$-tagging efficiency (60\%). Right: softened tight $b$-tagging efficiency (70\%). } 
\label{fig:ROC}
\end{figure}

The BDT tool is applied to the five exclusive analyses after event selection. Considering that $5b1 \ell$ and SS$2\ell$ represent probably the two most promising analyses at HL-LHC, we show their normalized event distributions w.r.t. BDT response in Figure~\ref{fig:ChannelHistograms}. The BDT method results in a wider separation between signal and backgrounds in the $5b1 \ell$ analysis, compared to the SS$2l$ one. Because suffering more from the combinatorial backgrounds due to the $b$-jet multiplicity in its final state, this channel gains more from the BDT method. 

Based on the BDT response of signal and backgrounds, the ROCs, the gain of statistical significance and the gain of $S/B$ are shown as a function of signal efficiency (or the threshold of BDT response) in Figure~\ref{fig:ROC}, for all of the five exclusive analyses. The statistical significances are optimized for some signal efficiency between $50\% - 90\%$ which improves the cut-based sensitivities by a factor $\sim 1.2 - 1.3$ for $5b1 \ell$ and $5b2 \ell$ and a factor $ \sim 1.1 - 1.2$ for the other three. Compared to the statistical significance, these analyses actually gain more from BDT in improving the $S/B$ values. For the signal efficiency with an optimized significance, the $S/B$ values are  improved by a factor of $\sim 1.5 - 3.5$. For example, the $S/B$ value in the analyses of $5b1 \ell$ and $5b2 \ell$ are raised from sub percent to precent level. An softened tight $b$-tagging efficiency may yield a better separation between signal and backgrounds, especially for the analyses of $5b1 \ell$ and $5b2 \ell$, and hence improve their sensitivities further (similar for the cut-based analyses). This can be seen by comparing the panels between the left and right columns in Figure~\ref{fig:ROC}.  But, in that case we need to simulate the impact of the backgrounds $tt+4c$, $2b2j$, $2c2j$, and $4j$ in a more solid way.

The optimized statistical significances and the correspondent $S/B$ values are summarized in Table~\ref{tab:cutflow1}. The $5b1 \ell$ and SS$2\ell$ analyses provide a sensitivity at HL-LHC more promising than the other ones, for the $tthh$(SM) measurement. If combined with multi-$\ell$ (in quadrature), the SS$2\ell$ analysis will result in a significance $\sim 0.6\sigma$, comparable to that of the $5b1 \ell$ one.   
A combination of all allows measuring the $tthh$(SM) production with a significance $\sim 0.9\sigma$.

\subsection{Kinematics and Sensitivities of $\kappa$ and $c_t$}

\begin{figure}[th]
\centering
\includegraphics[scale=0.2]{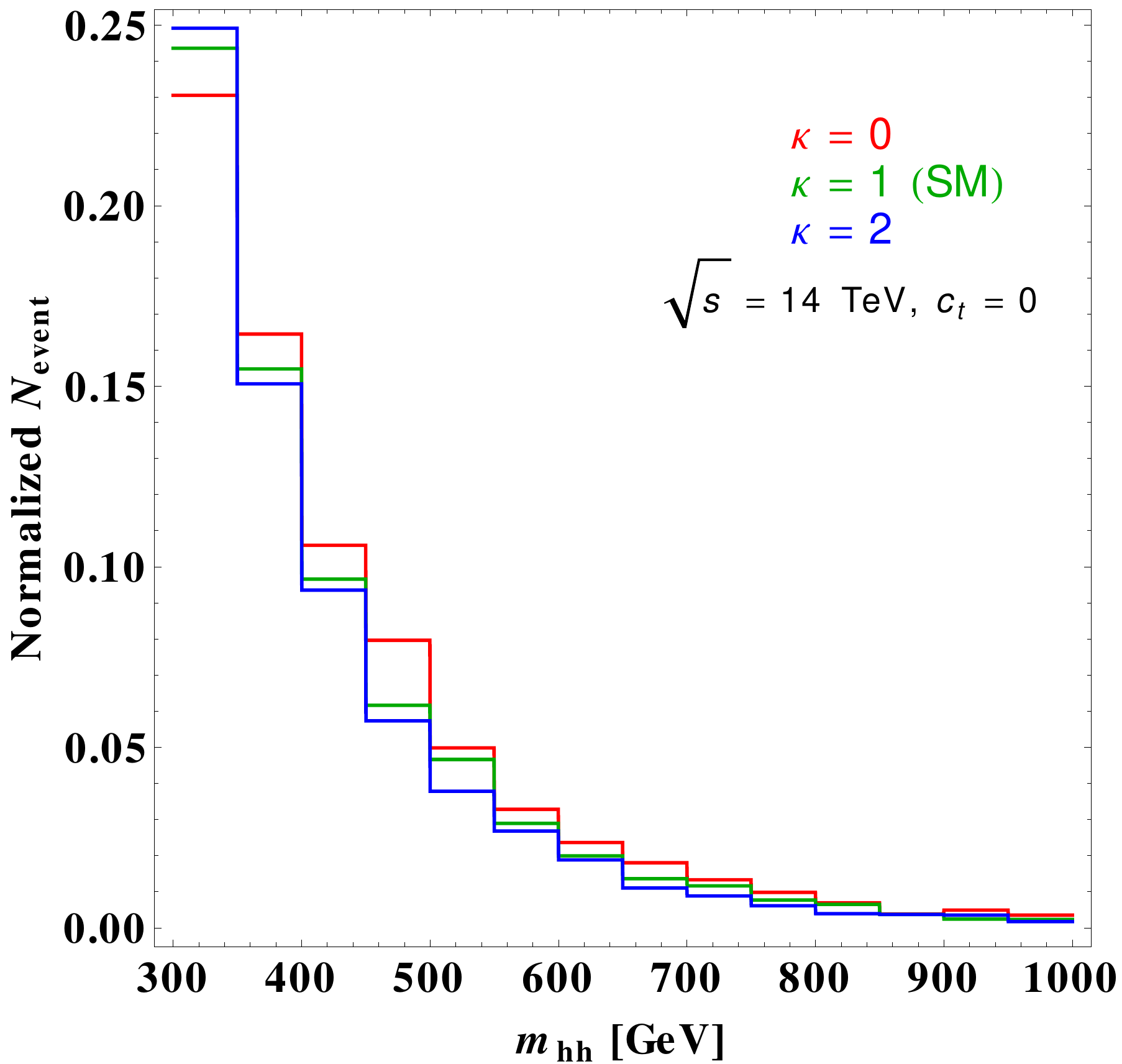}
\includegraphics[scale=0.2]{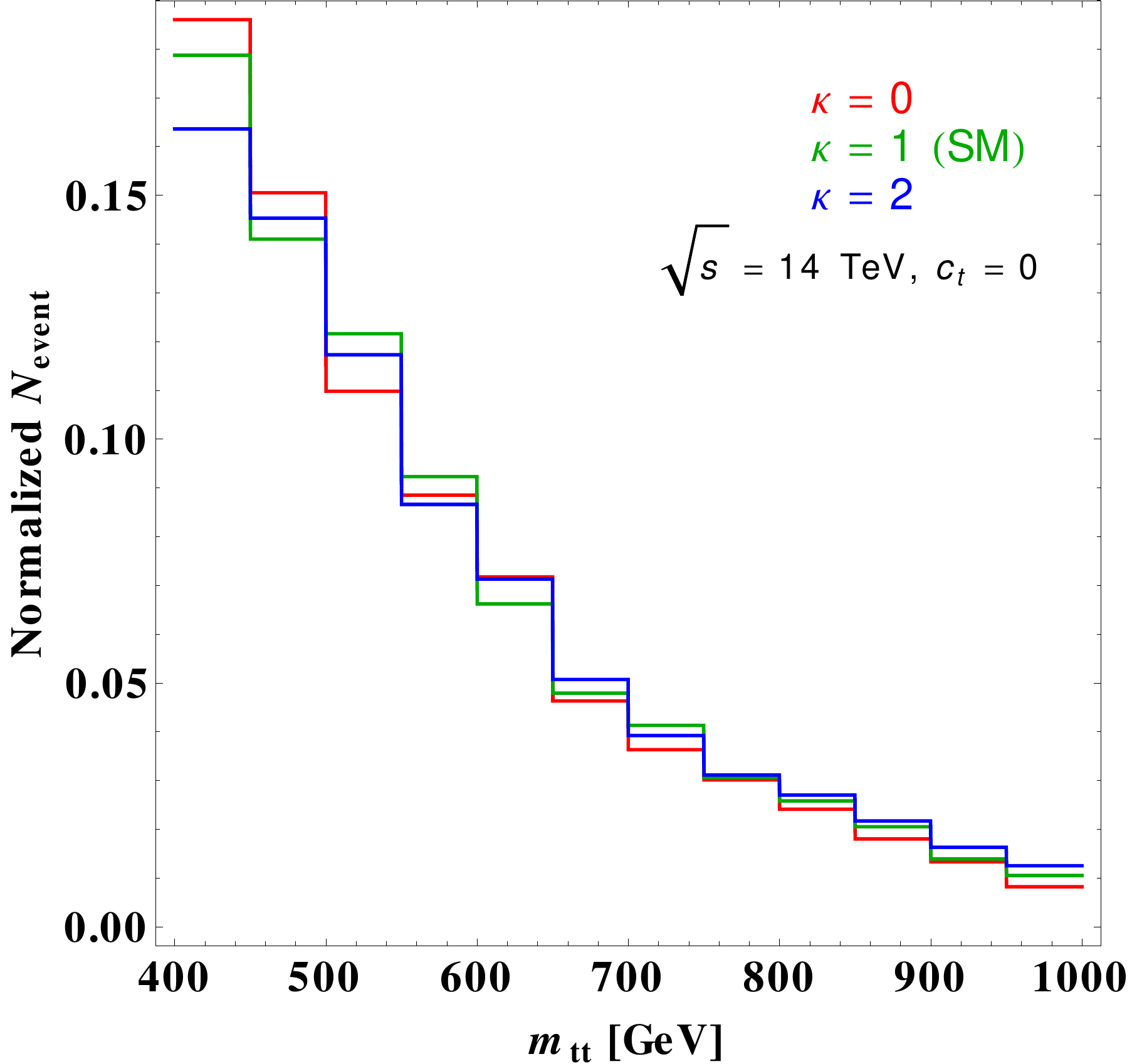}
\includegraphics[scale=0.2]{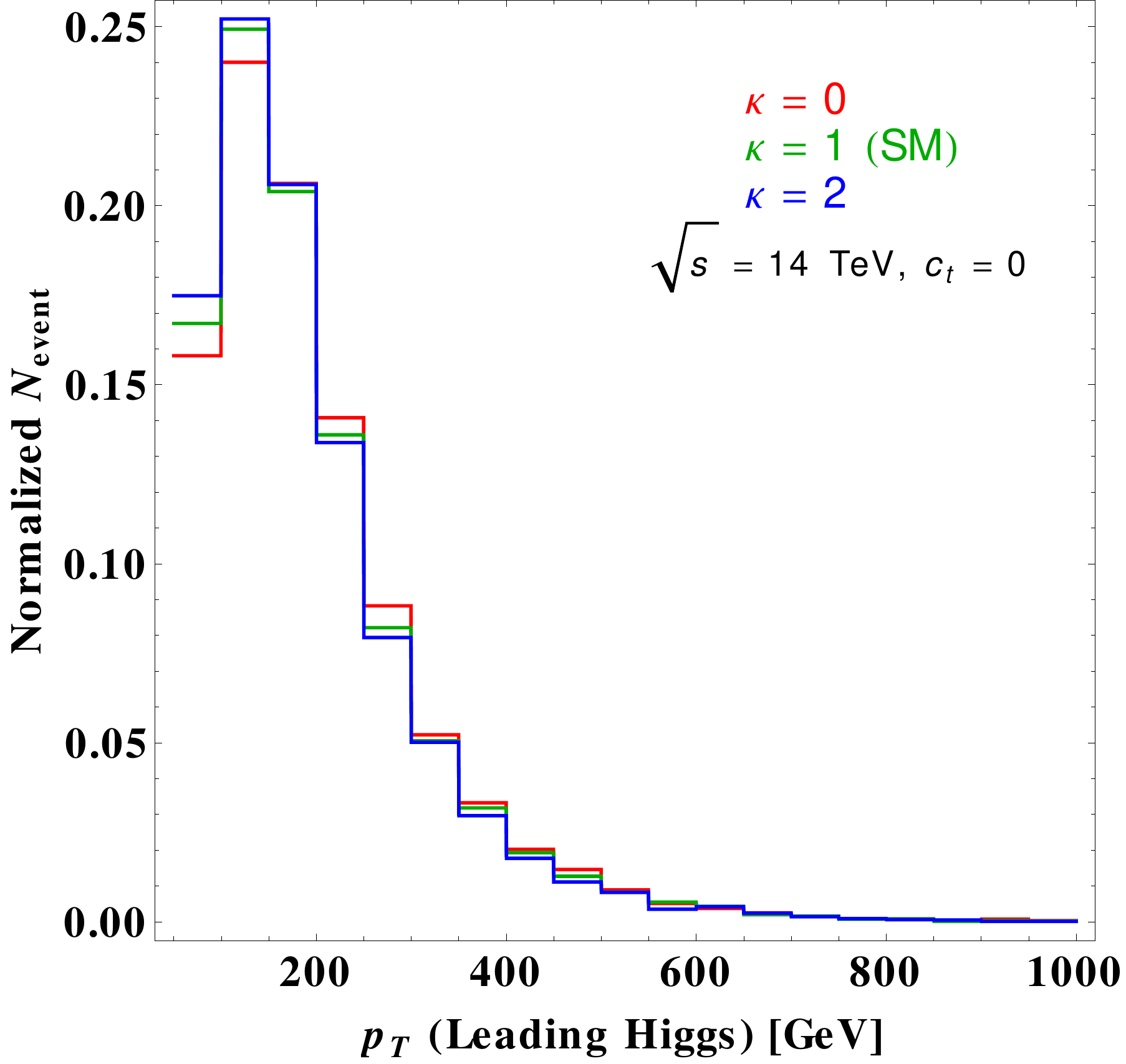}
\includegraphics[scale=0.2]{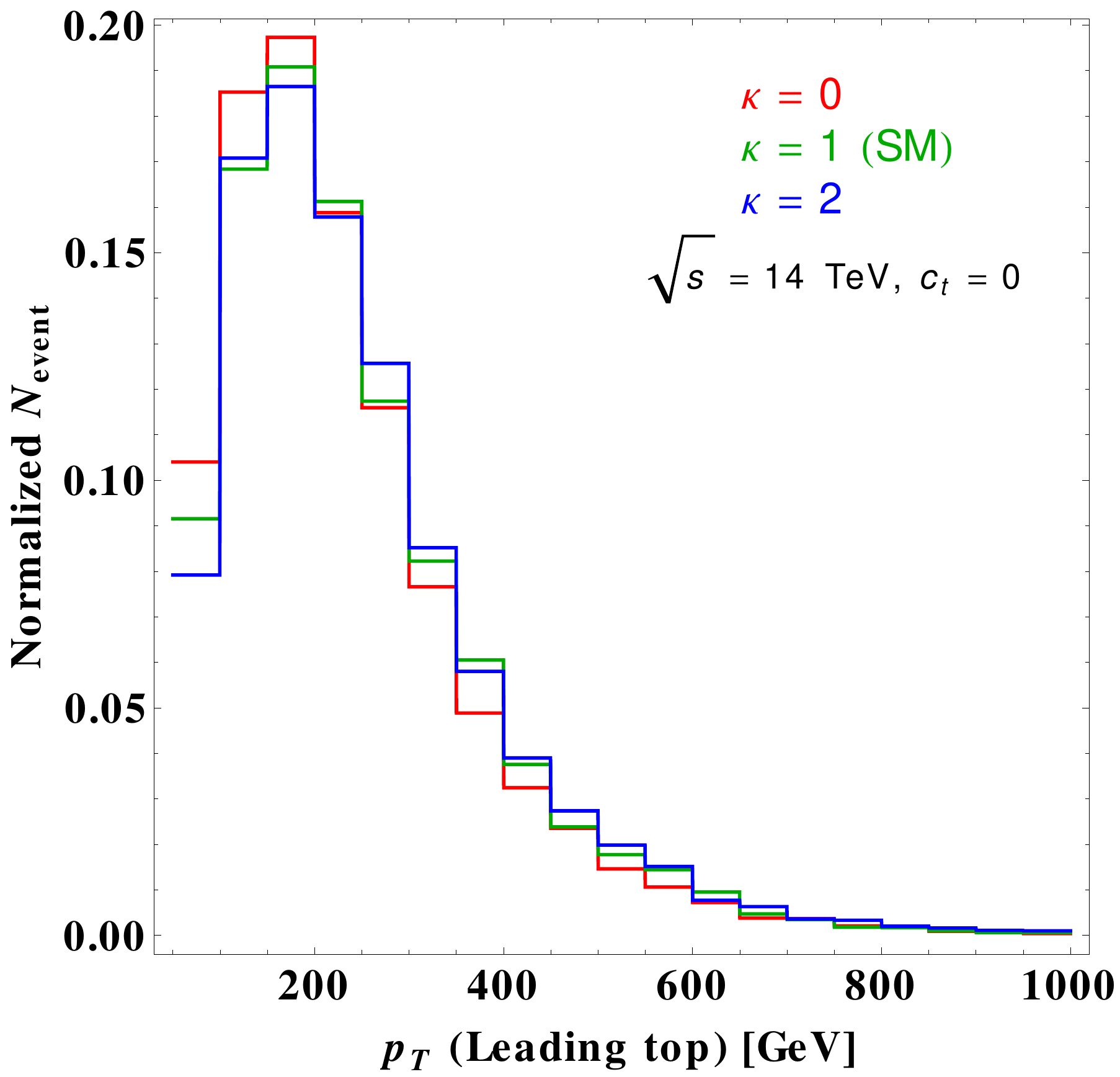}
\includegraphics[scale=0.2]{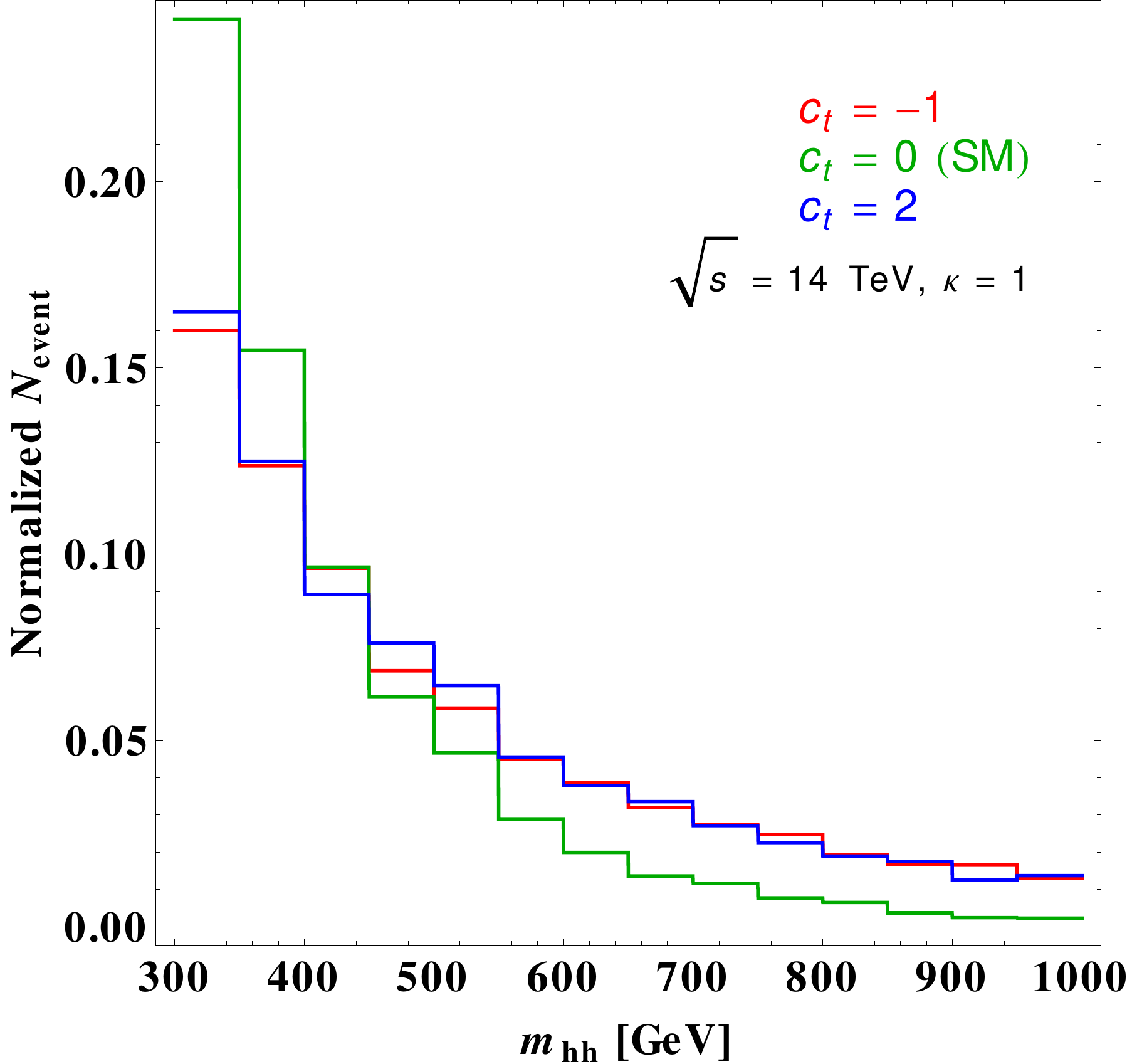}
\includegraphics[scale=0.2]{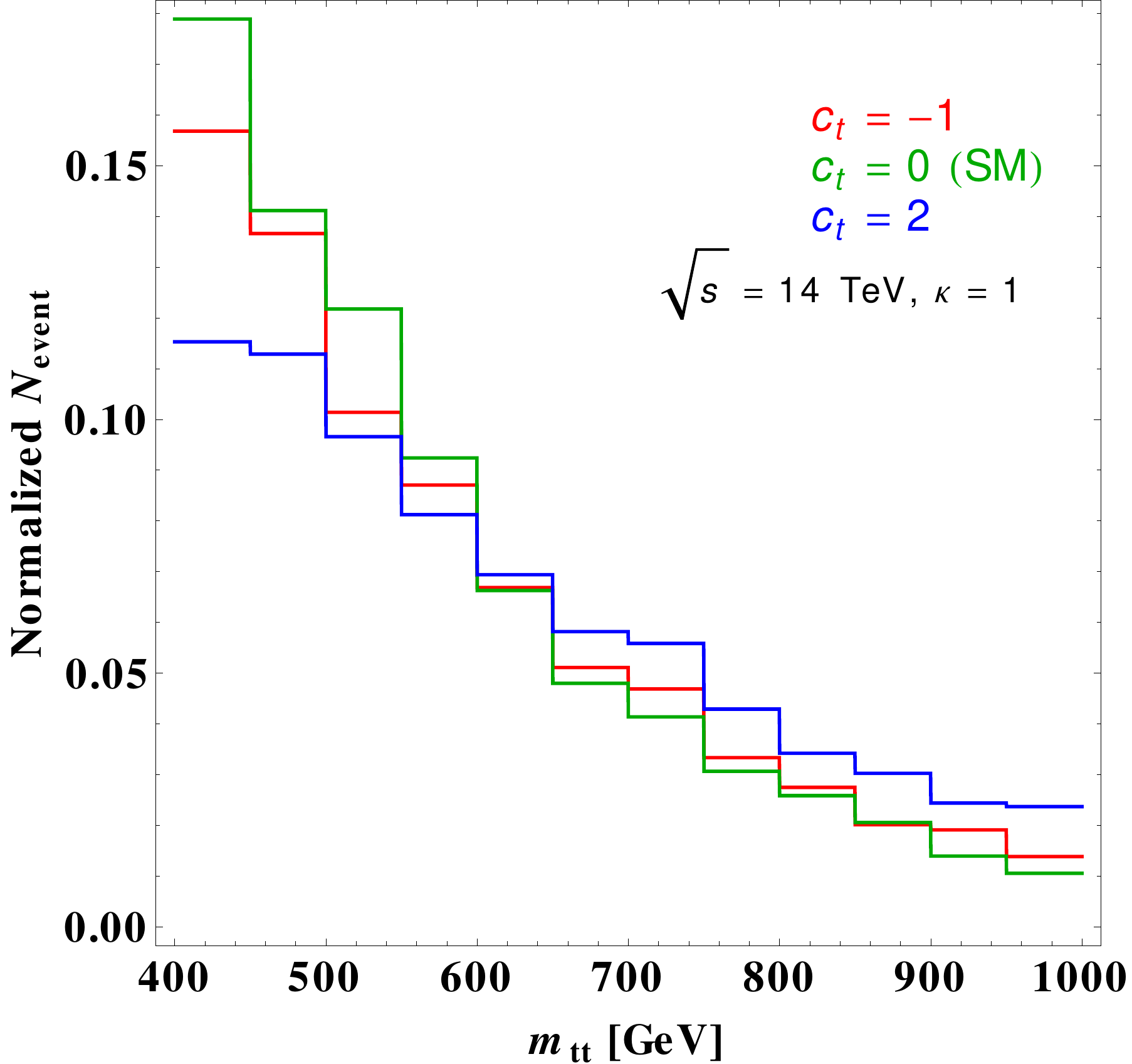}
\includegraphics[scale=0.2]{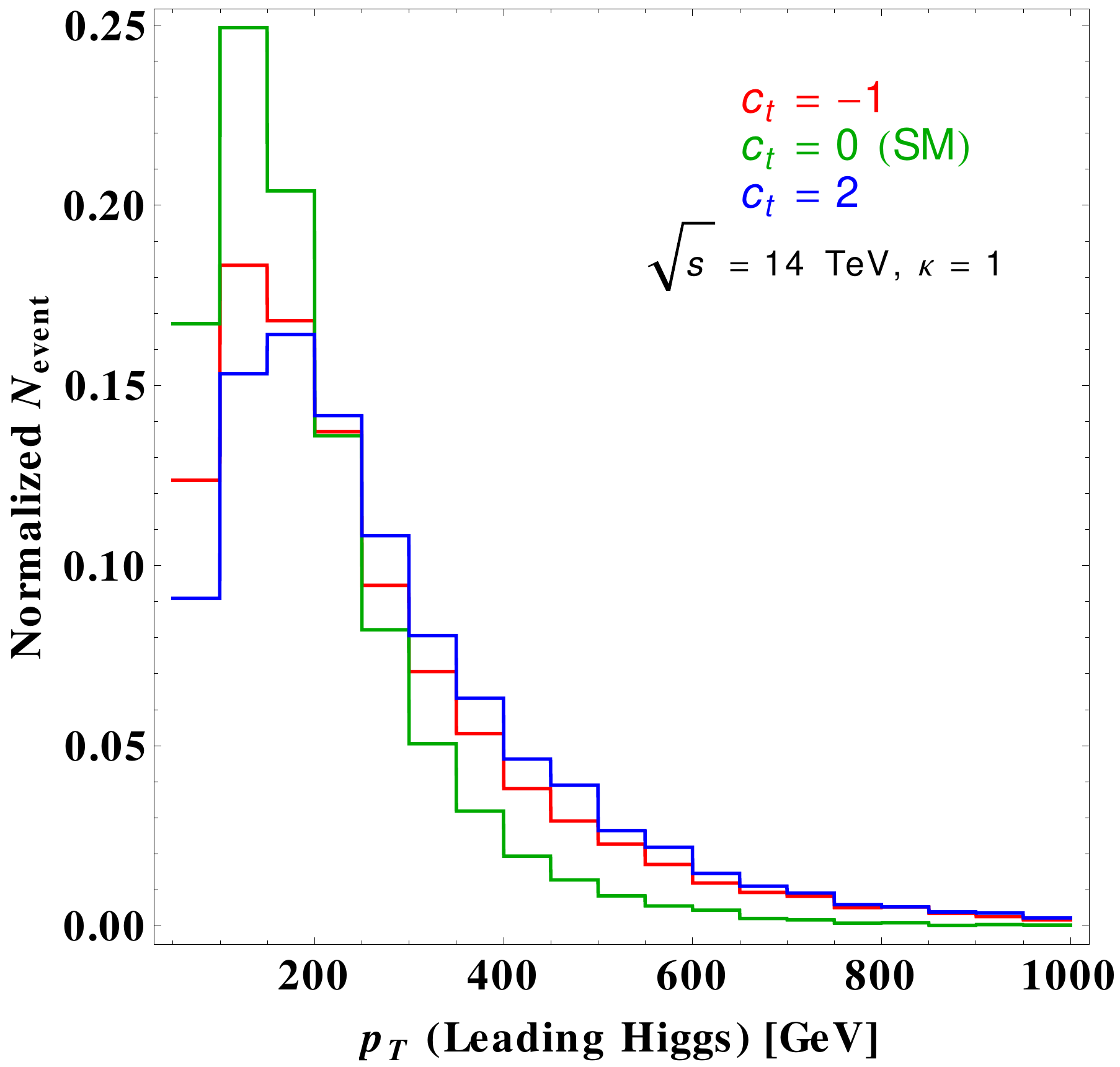}
\includegraphics[scale=0.2]{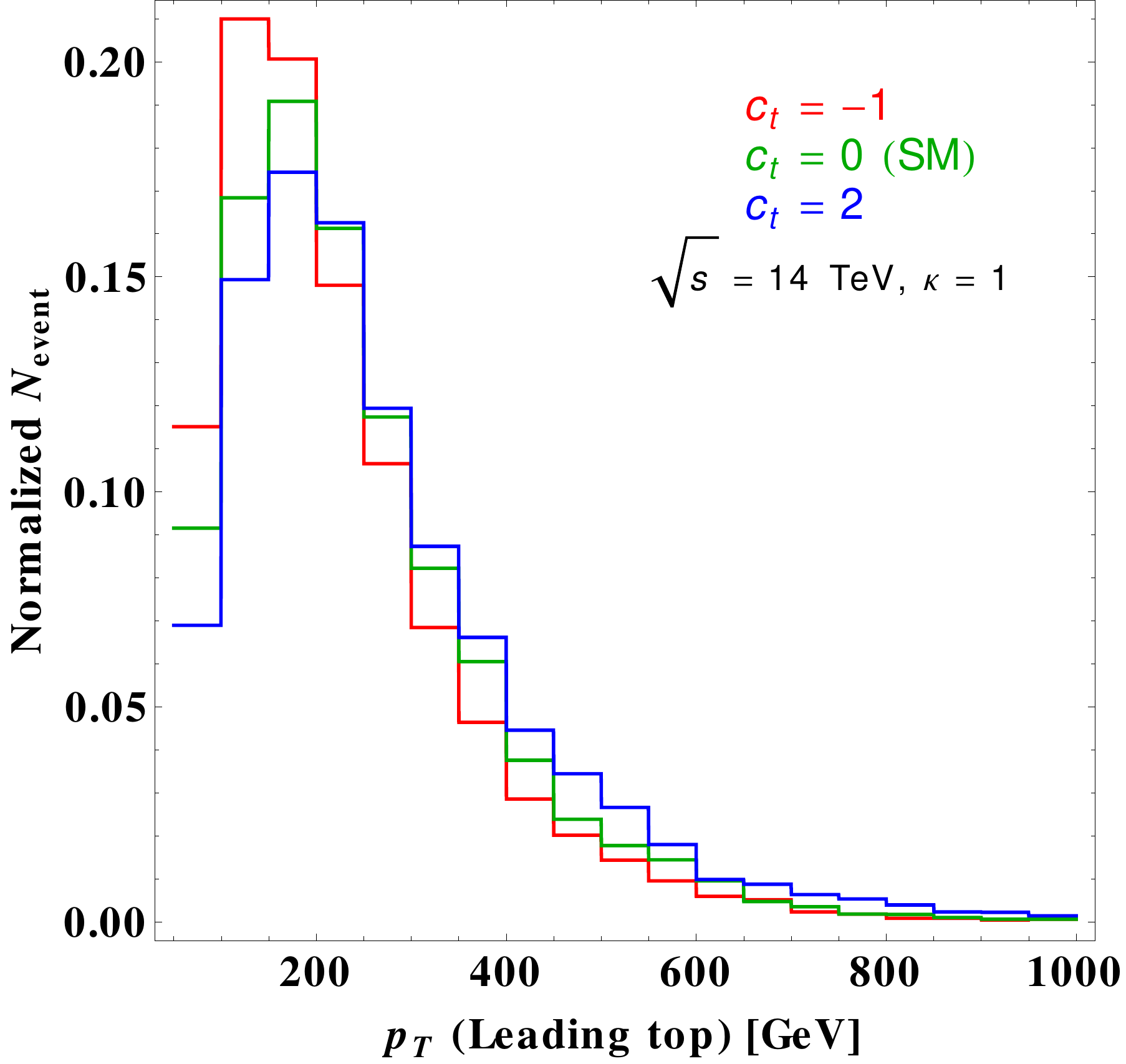}
\caption{Normalized kinematic distributions in terms of $\kappa$ (top) and $c_t$ (bottom) at parton level, at HL-LHC. $c_t$ and $\kappa$ are fixed to their SM values in  top and bottom panels, respectively.}
\label{fig:kinematic}
\end{figure}

As introduced in Section~\ref{sec:intro}, the $tthh$ production can serve as a probe for the anomalous Higgs self-interaction and the $tthh$ contact interaction~\cite{Contino:2012xk,Chen:2014xra}. In principle we can apply the BDT model which is trained for the $tthh$(SM) measurement, to extract the relevant sensitivity information. But, the variation of  the Higgs self-coupling and the $tthh$ contact interaction may bring new kinematic features to assist the separation between signal and backgrounds. So it will be useful to make a study on such features.

\begin{figure}[th]
\centering
\includegraphics[scale=0.35]{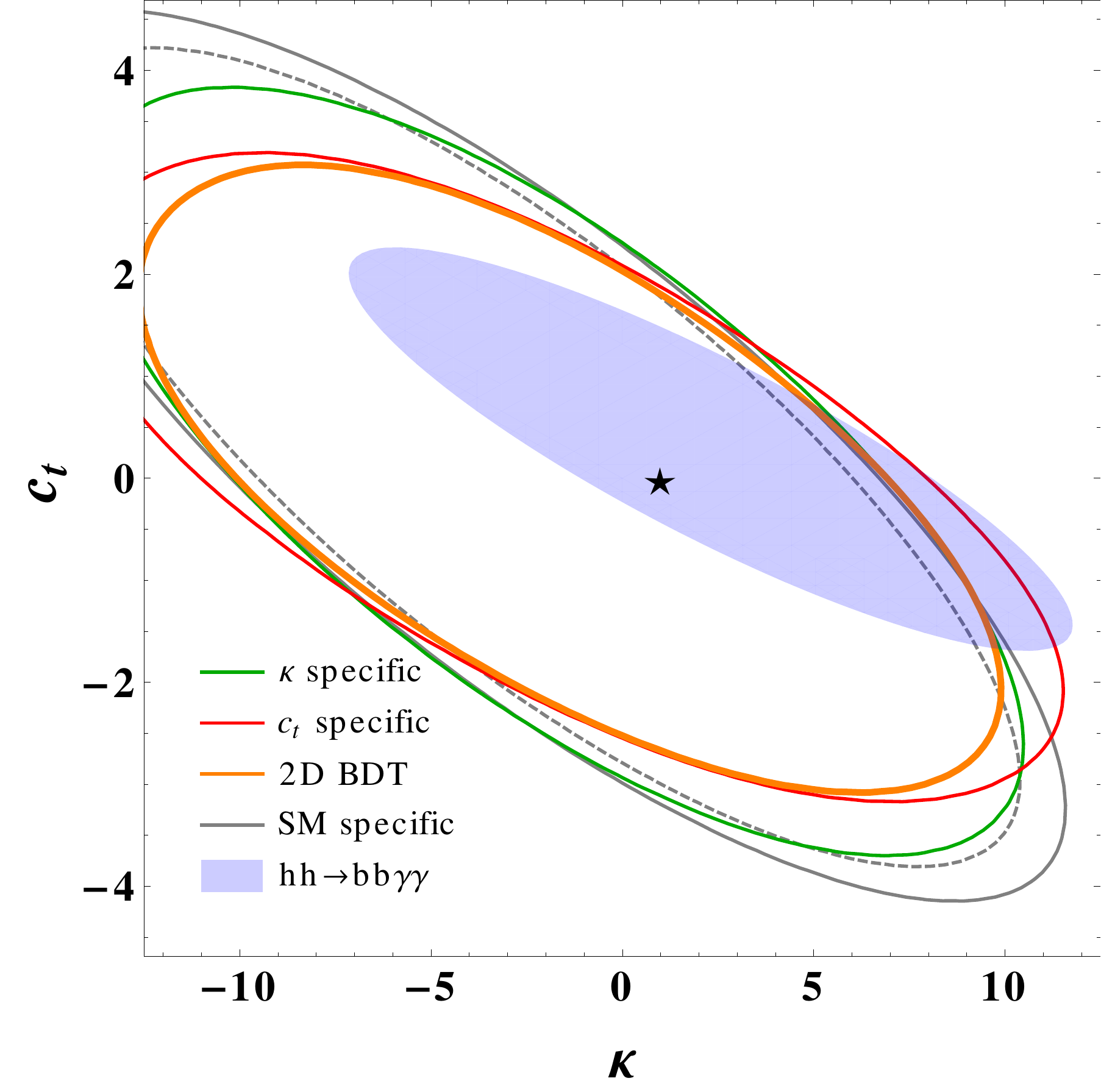}
\includegraphics[scale=0.35]{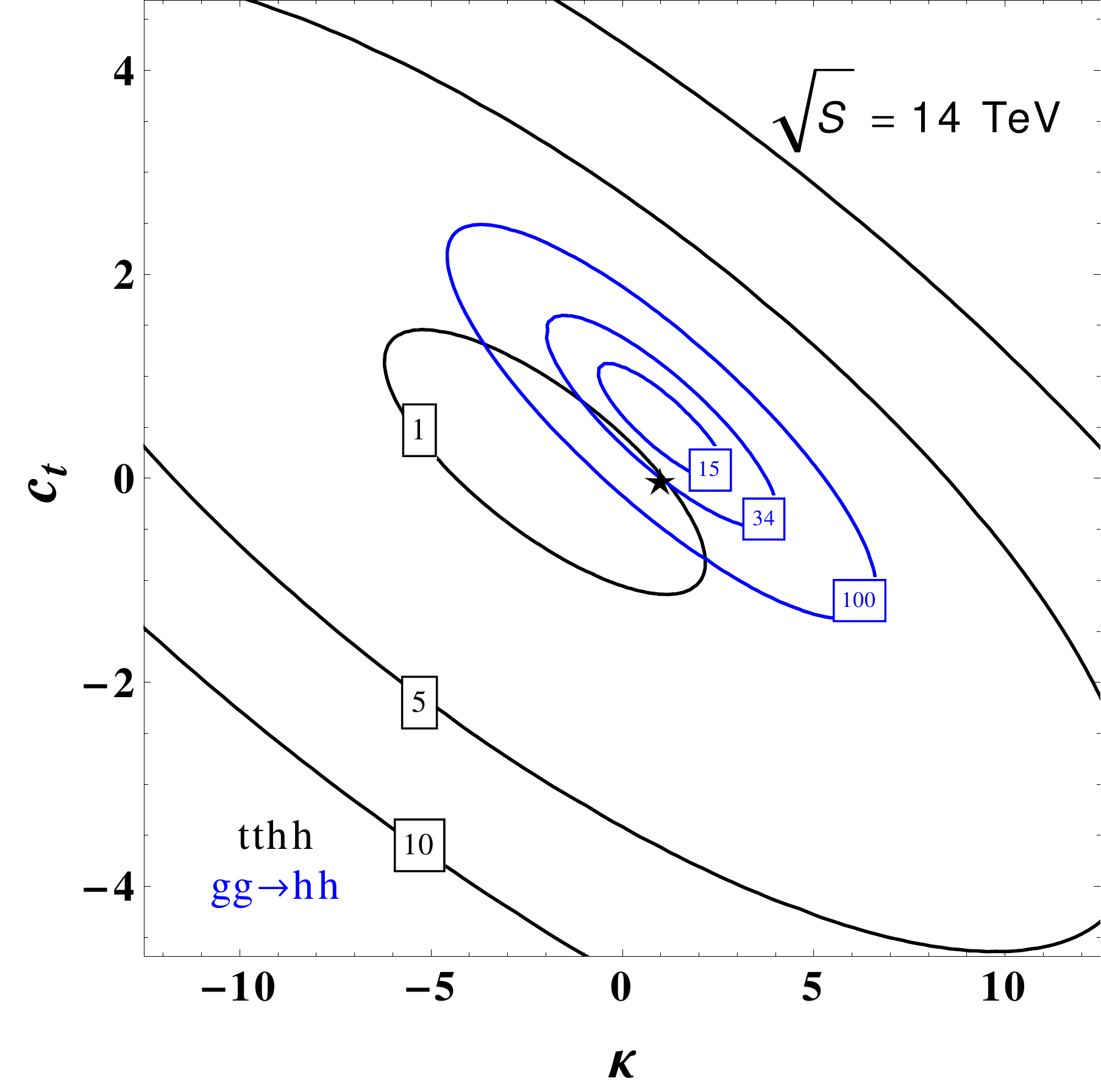}
\caption{Exclusion contours (left), and $\sigma(tthh)_{14}$ (fb), $\sigma(gg\to hh)_{14}$ (fb) contours (right) in terms of $\kappa$ and $c_t$ at HL-LHC. The exclusion limits are defined against the hypothesis of background (including $tthh$(SM)) + signal (deviation from $\{\kappa, c_t\} = \{1,0\}$).}
\label{fig:exconts}
\end{figure}

In Figure~\ref{fig:kinematic}, we show the normalized kinematic distributions of the $tthh$ events at parton level, w.r.t. varied $\kappa$ and $c_t$ values (for discussions on the impact of the six-dimensional operator $\partial_\mu (H^\dagger H) \partial^\mu (H^\dagger H)$ for the $tthh$ kinematics at parton level, see~\cite{He:2015spf}). It is easy to see that, as $\kappa$ increases, the $tt$ invariant mass tends to be larger, while the $hh$ invariant mass tends to be smaller. But, the dependence of $p_T$ on the $\kappa$ value is relatively weak for the leading Higgs boson or top quark. Differently, a deviation of $c_t$ from its SM value, {\it i.e.}, $c_t =0$ tends to yield a larger invariant mass for both $hh$ and $tt$ pairs in the events. The leading Higgs boson also tends to be harder. These features are strongly correlated with the $y^2\kappa^2$- and $c_t^2$-mediated dynamics in the $tthh$ production, which will be manifested later in Figure~\ref{fig:kin2} and the relevant discussions in the text.  To utilize these features, therefore, we train two BDT models using the specific samples of $y^2\kappa^2$ and $c_t^2$ respectively (more information on the sample generation can be found in subsection~\ref{sec:simulation}). Then the sensitivity to probe for $\kappa$ and $c_t$ via the $tthh$ production is analyzed using the 2D BDT responses for optimization. 

A comparison of exclusion limits at HL-LHC, which are generated using different BDT models, is presented in left panel of Figure~\ref{fig:exconts}. In this panel, the blue region represents the sensitivity reach of a reference analysis of $gg\to hh \to bb \gamma\gamma$ pursued by the ATLAS collaboration~\cite{ATL-PHYS-PUB-2018-053}, which remains not excluded at HL-LHC. This analysis was taken in the limit of $c_t=0$. To project its results to the $\kappa - c_t$ plane, we use the relation~\cite{Azatov:2015oxa}  
\begin{equation}
\frac{\sigma(gg\to hh \to bb\gamma \gamma)_{14}}{\sigma(gg\to hh \to bb\gamma \gamma)_{14}^{\rm SM}} = 1.70 -0.82 \kappa + 0.12 \kappa^2  - 3.79 c_t + 0.98 c_t \kappa +2.68 c_t^2~,
\label{eq:gg1}
\end{equation}
by requiring the exclusion contour to reduce to the ATLAS results in the limit of $c_t=0$. Note, this relation was generated with some basic cuts being applied, for the two $b$-jets and the two photons in the signal events and for the Higgs invariant mass reconstructed from them~\cite{Azatov:2015oxa}. Though some imprecision can be caused by the mismatch between the phase spaces involved in these two analyses, we would expect that it will not qualitatively change our conclusions. So we will tolerate this imprecision in the discussions. In the blue region, the central part can be probed for with a relatively higher sensitivity, compared to the iso-significance elliptical belt near the edge. This is related to the deviation of the signal rate from its SM prediction, to a large extent. In the central part, the signal rate is largely suppressed (also see the $\sigma(gg\to hh)_{14}$ contours in the right panel of Figure~\ref{fig:exconts}) due to destructive interference, as is manifested by the minus sign of the $\kappa$ and $c_t$ linear terms in Eq. (\ref{eq:gg1}). As for the elliptical belt near the edge, the predicted signal rate is degenerated or approximately degenerated with the SM one. 

The colored curves in this panel represent the exclusion limits generated using different BDT models. For each specific BDT model, we generate one set of signal efficiency and background rejection $\{\epsilon_{\rm sig}^{ij}, \epsilon_{\rm bg}^{i}\}$. Here $i$ runs over the five exclusive analyses, and $j$ runs over the six samples defined by $\{y^4, y^2\kappa^2, c_{t}^2, y^3\kappa, y^2c_{t}, y\kappa c_{t}\}$. Different from $\epsilon_{\rm sig}^{ij}$, $\epsilon_{\rm bg}^{i}$ is fixed in favor of some specific signal sample (e.g., the $y^2\kappa^2$ sample or the $c_t^2$ sample), and hence be universal for all of the six samples in each analysis.  With this information, we will be able to calculate the sensitivity for any given values of $\{\kappa, c_t\}$, by reweighting the relative contributions of these six samples, using the LO relation at 14 TeV (generated with MadGraph~5~\cite{Alwall:2011uj}): 
\begin{eqnarray}
\frac{\sigma(tthh)_{14}}{\sigma(tthh)_{14}^{\rm SM}} &=&0.82+0.14\kappa + 0.04 \kappa^2 + 0.28 c_t + 0.21 \kappa c_t + 0.44 c_t^2 \ . 
\label{eq:sig14}
\end{eqnarray}
Here $\sigma(tthh)_{14}^{\rm SM}=0.99$ fb is the $tthh$(SM) cross sections at 14 TeV. The $\sigma(tthh)_{14}$ contours in terms of $\kappa$ and $c_t$ are presented in right panel of Figure~\ref{fig:exconts}. Obviously, their absolute gradient is larger along the line from bottom-left to top-right, compared to its orthogonal direction. This is because the quadratic terms of $\kappa$ and $c_t$ in Eq. (\ref{eq:sig14}) contribute to $\sigma(tthh)_{14}$ in a topping-up way along this line. Differently, a cancellation occurs between the $y^2\kappa^2$, $c_t^2$ terms and the $\kappa c_t$ term in the orthogonal direction of this line. 

In the left panel of Figure~\ref{fig:exconts}, the solid (dashed) grey contour is set by the specific BDT model which is trained for the SM $tthh$ analyses with a tight (softened tight) $b$-tagging efficiency of 60\% (70\%). The exclusion contours based on the $t^2\kappa^2$- and $c_t^2$-specific  BDT models are presented in green and red in the figure, respectively. The $t^2\kappa^2$-specific BDT model can also improve the sensitivity to $c_t^2$. This is because the $t^2\kappa^2$ and $c_t^2$ samples share similar kinematics on $m_{t\bar{t}}$. The orange contour represents the sensitivity resulting from the 2D analysis, based on the two specific BDT models. The impact of the $\sigma(tthh)_{14}$ magnitude and the $\kappa$- and $c_t$-induced kinematics for the analysis sensitivity can be easily seen by comparing the contours in Figure~\ref{fig:exconts}: the $\sigma(tthh)_{14}$ magnitude sets up the orientation of the exclusion contours, whereas the $\kappa$- and $c_t$-induced kinematics further improves them. These contours all exclude the right edge of the blue area, despite of their difference in extent, and partially breaks the degeneracy which is usually thought to exist in the $gg\to hh \to bb \gamma\gamma$ analysis at HL-LHC~\footnote {A combination of the $gg\to hh \to bb \gamma\gamma$ analysis and the other ones of $gg\to hh$ may not completely remove this degeneracy at HL-LHC~\cite{ATL-PHYS-PUB-2018-053}, though it can be achieved at 27 TeV or 100 TeV using the kinematic difference in the degenerated parameter regions.}.

\subsection{Projected Sensitivities at Future Hadron Colliders}

\begin{figure}[th]
\centering
\includegraphics[scale=0.35]{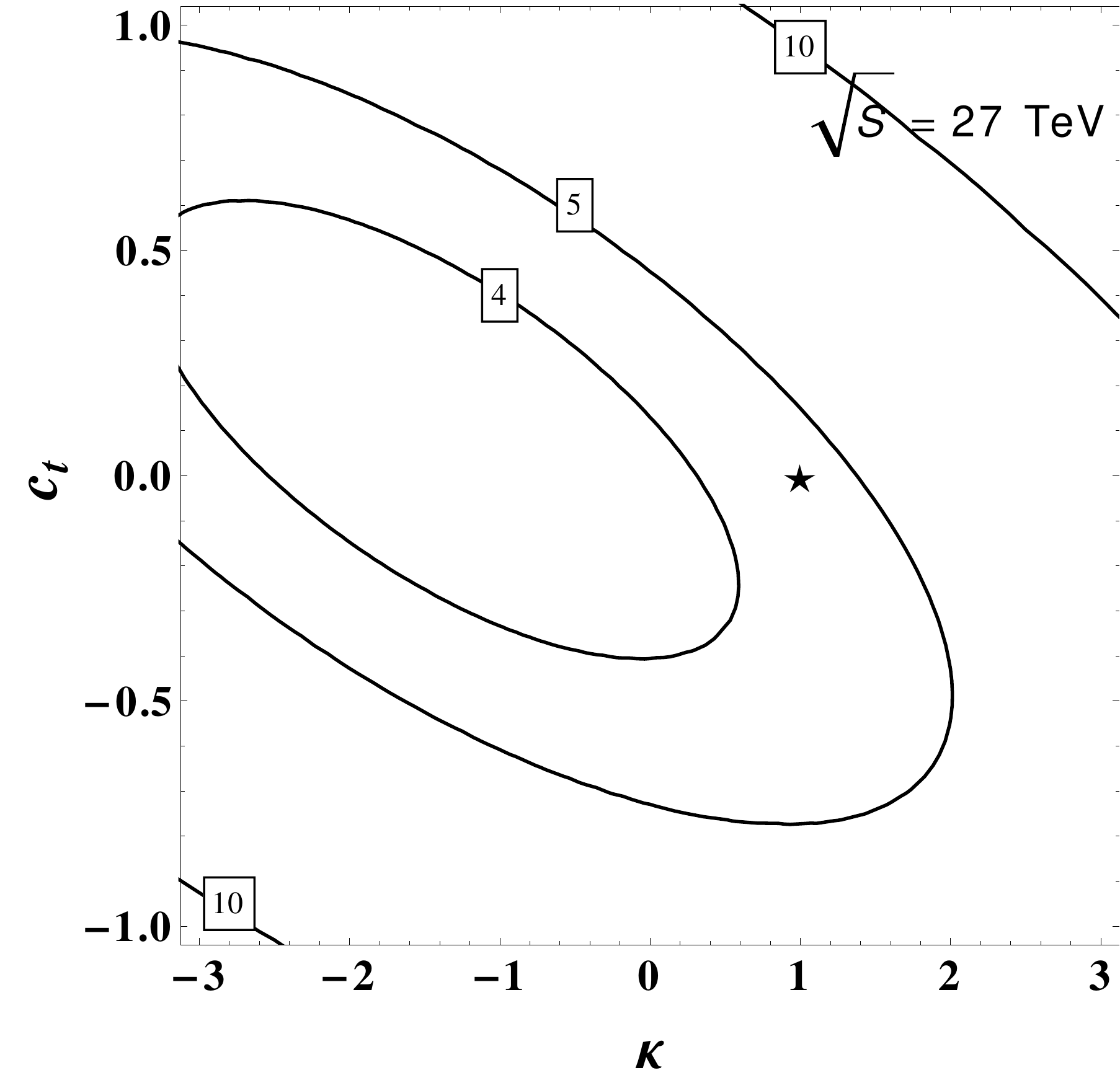} \ \ \ \ 
\includegraphics[scale=0.35]{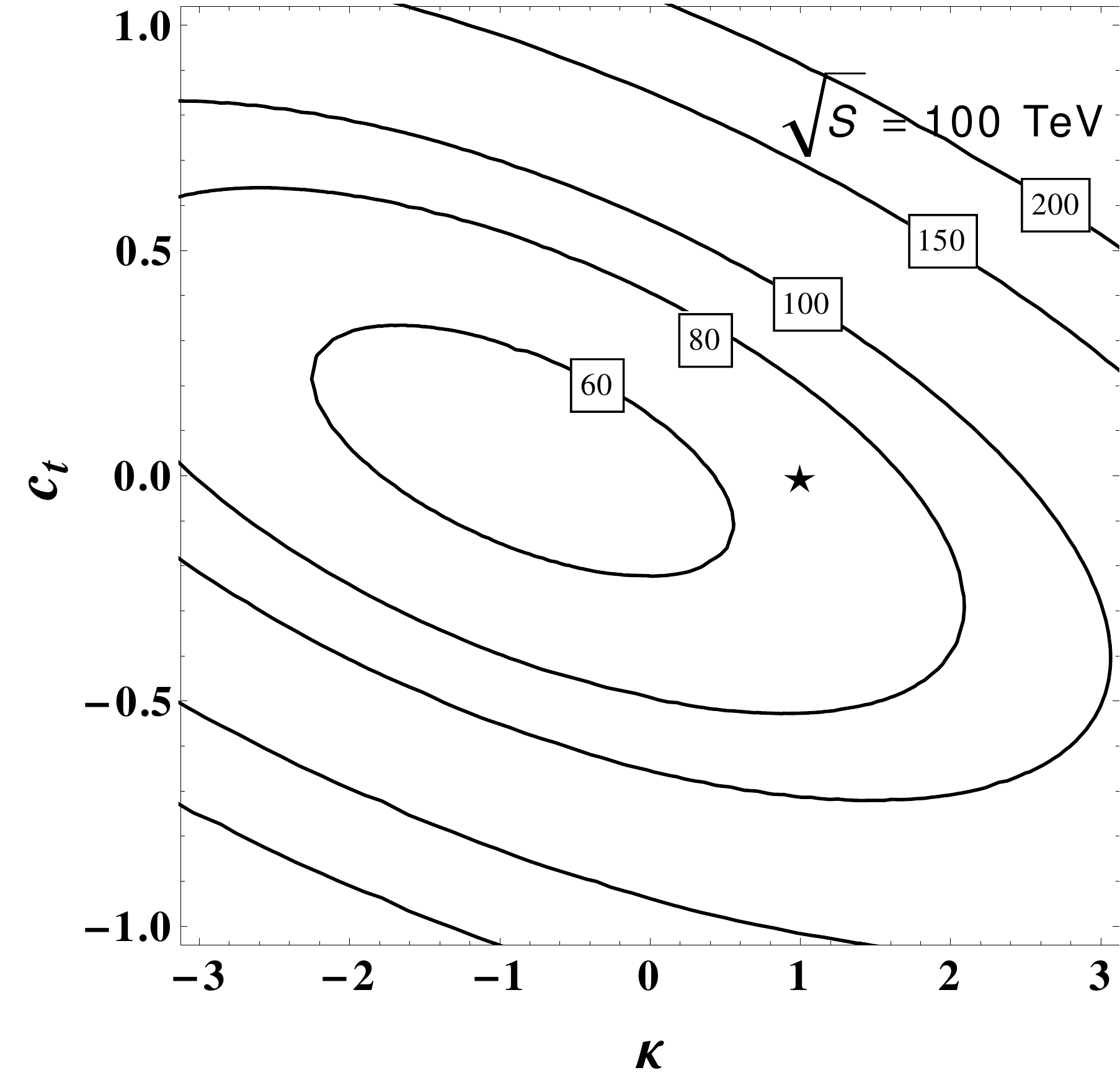}
\caption{$\sigma(tthh)_{27}$ (fb) (left) and $\sigma(tthh)_{100}$ (fb) (right) contours in terms of $\kappa$ and $c_t$. 
}
\label{fig:sigma27+100}
\end{figure}

The study on the $tthh$ physics will benefit a lot from future hadron colliders, such as the increased signal rate, the enhanced boost of signal kinematics, and even a larger luminosity. Pursuing comprehensive analyses on this is beyond the scope of this article. Instead, we will project the $tthh$ sensitivities at HL-LHC  to 27 TeV and 100 TeV, to qualitatively measure their potential. 

The projection of the $tthh$(SM) measurement sensitivities is straightforward. It gives a significance of $3.1\sigma$ and $14.3\sigma$ at 27 TeV and 100 TeV, respectively, based on the combined BDT analyses. To project the exclusion limits of 2D BDT at the $\kappa - c_t$ plane, we simulate the LO $tthh$ production rate at 27 TeV and 100 TeV as a function of $\kappa$ and $c_t$  
\begin{eqnarray}
\frac{\sigma(tthh)_{27}}{\sigma(tthh)_{27}^{\rm SM}} &=& 0.83+0.12 \kappa + 0.05 \kappa^2 + 0.22 c_t + 0.27 \kappa c_t + 0.78 c_t^2  \ , \\ 
\frac{\sigma(tthh)_{100}}{\sigma(tthh)_{100}^{\rm SM}} &=& 0.84+0.07 \kappa + 0.09 \kappa^2 + 0.15 c_t + 0.41 \kappa c_t + 1.73 c_t^2  \ ,
\label{eq:sig27100}
\end{eqnarray}
using MadGraph 5~\cite{Alwall:2011uj}. Here $\sigma(tthh)_{27}^{\rm SM}=4.58$ fb and  $\sigma(tthh)_{100}^{\rm SM}=67.41$ fb are the $tthh$(SM) cross sections at 27 TeV and 100 TeV, respectively.  The contours of $\sigma(tthh)_{27}$ and $\sigma(tthh)_{100}$ are presented in Figure~\ref{fig:sigma27+100}. To project the 100 TeV sensitivity of the reference analysis, {\it i.e.}, $gg\to hh\to bb \gamma\gamma$, to  the $\kappa-c_t$ plane, we will use the relation~\cite{Azatov:2015oxa} 
\begin{equation}
\frac{\sigma(gg \to hh \to bb\gamma \gamma)_{100}}{\sigma(gg\to hh \to bb\gamma \gamma)_{100}^{\rm SM}} = 1.59 -0.68 \kappa + 0.09 \kappa^2  - 3.83 c_t + 0.92 c_t \kappa +3.20 c_t^2~.
\label{eq:gg2}
\end{equation}
The relation at 27 TeV was not provided in~\cite{Azatov:2015oxa}, and is hence generated by interpolating the ones at 14 TeV and 100 TeV. Again we will tolerate the imprecision caused by using these relations for the analysis of $gg\to hh\to bb \gamma\gamma$. 

\begin{figure}[th]
\centering
\includegraphics[scale=0.35]{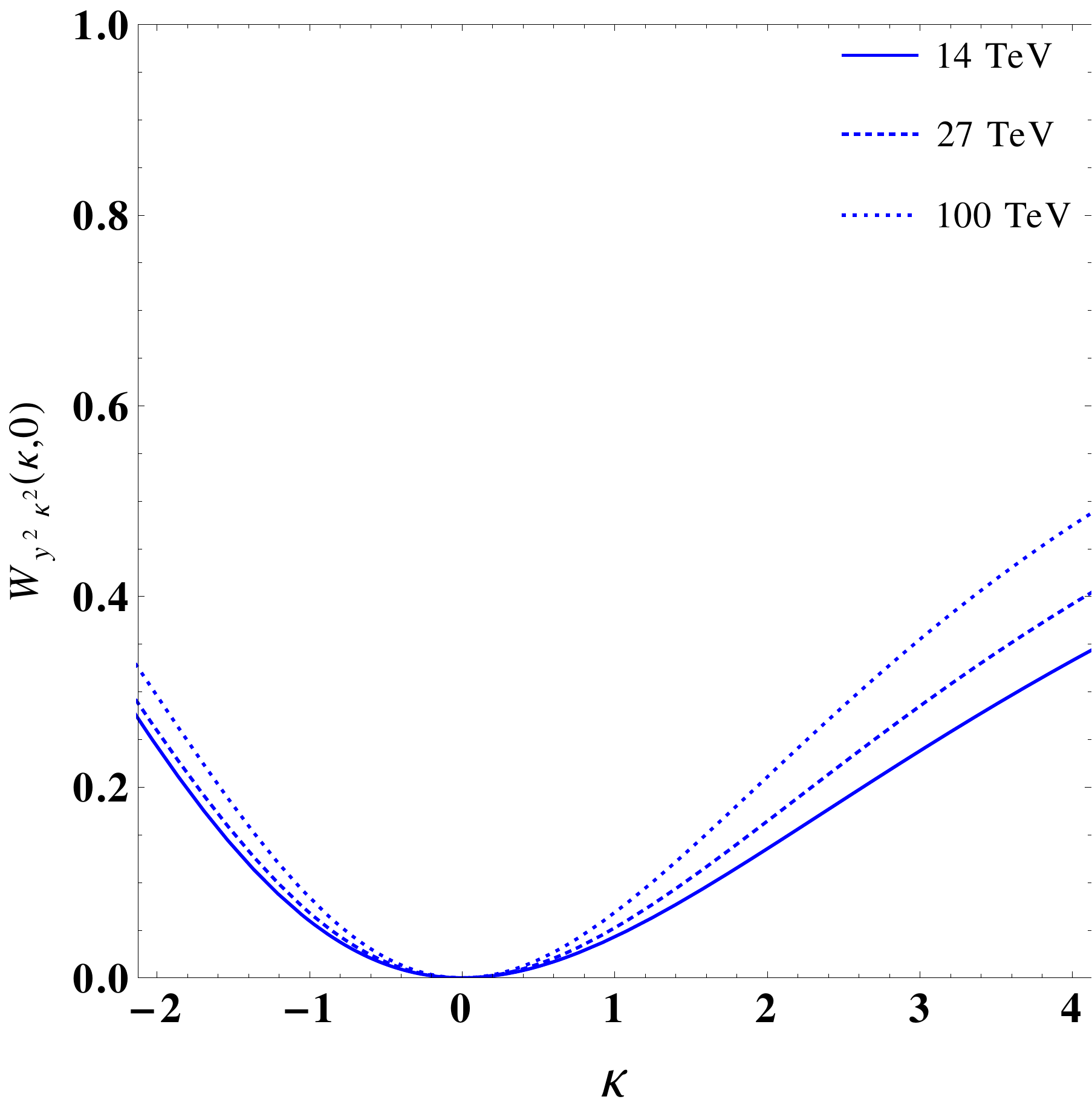} \ \ \ \ 
\includegraphics[scale=0.35]{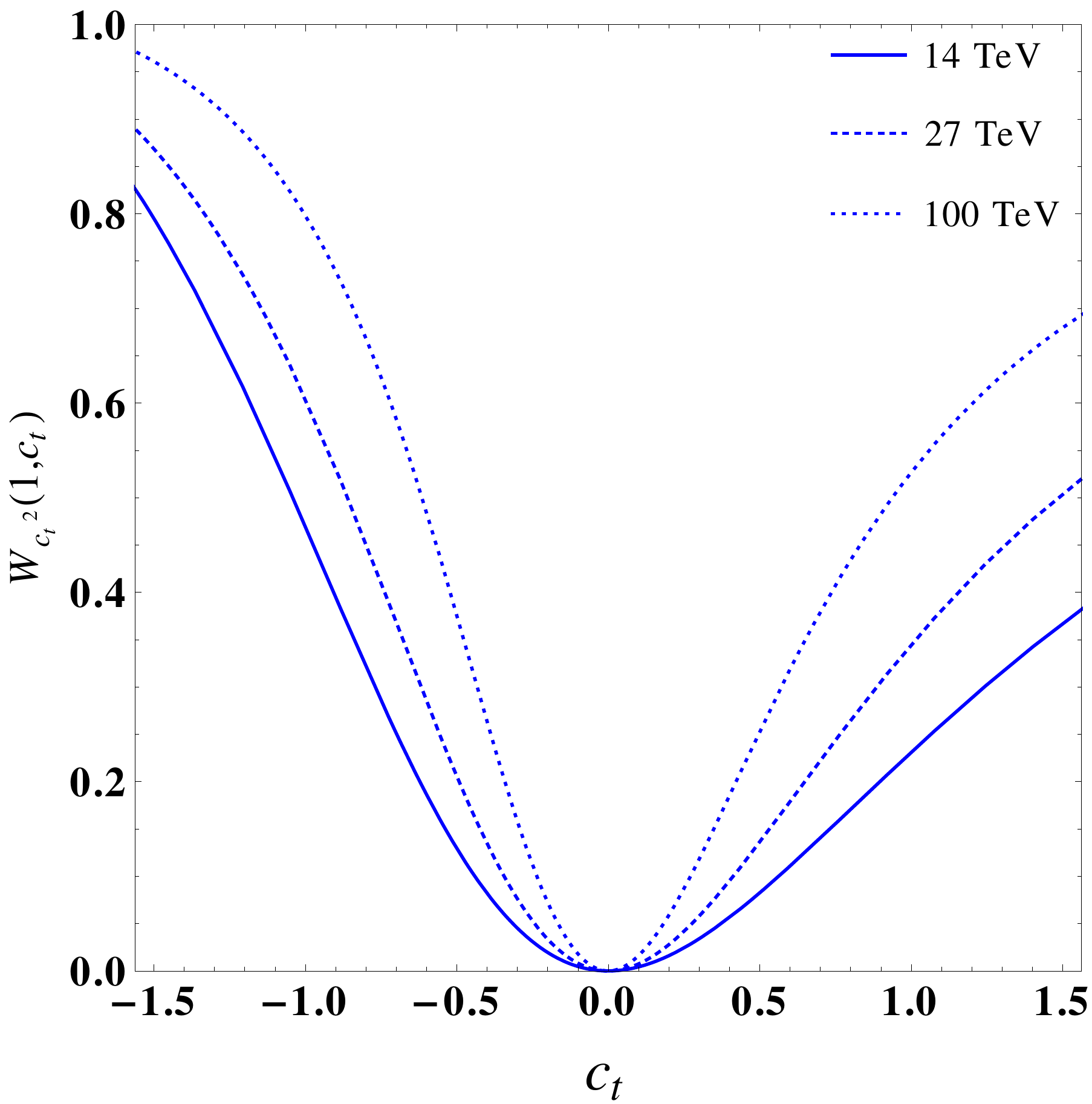}
\caption{The weights $W_{y^2\kappa^2}(\kappa, 0)$ and $W_{c_t^2}(1,c_t)$ at 14 TeV, 27 TeV and 100 TeV. 
}
\label{fig:kin1}
\end{figure}

\begin{figure}[th]
\centering
\includegraphics[scale=0.2]{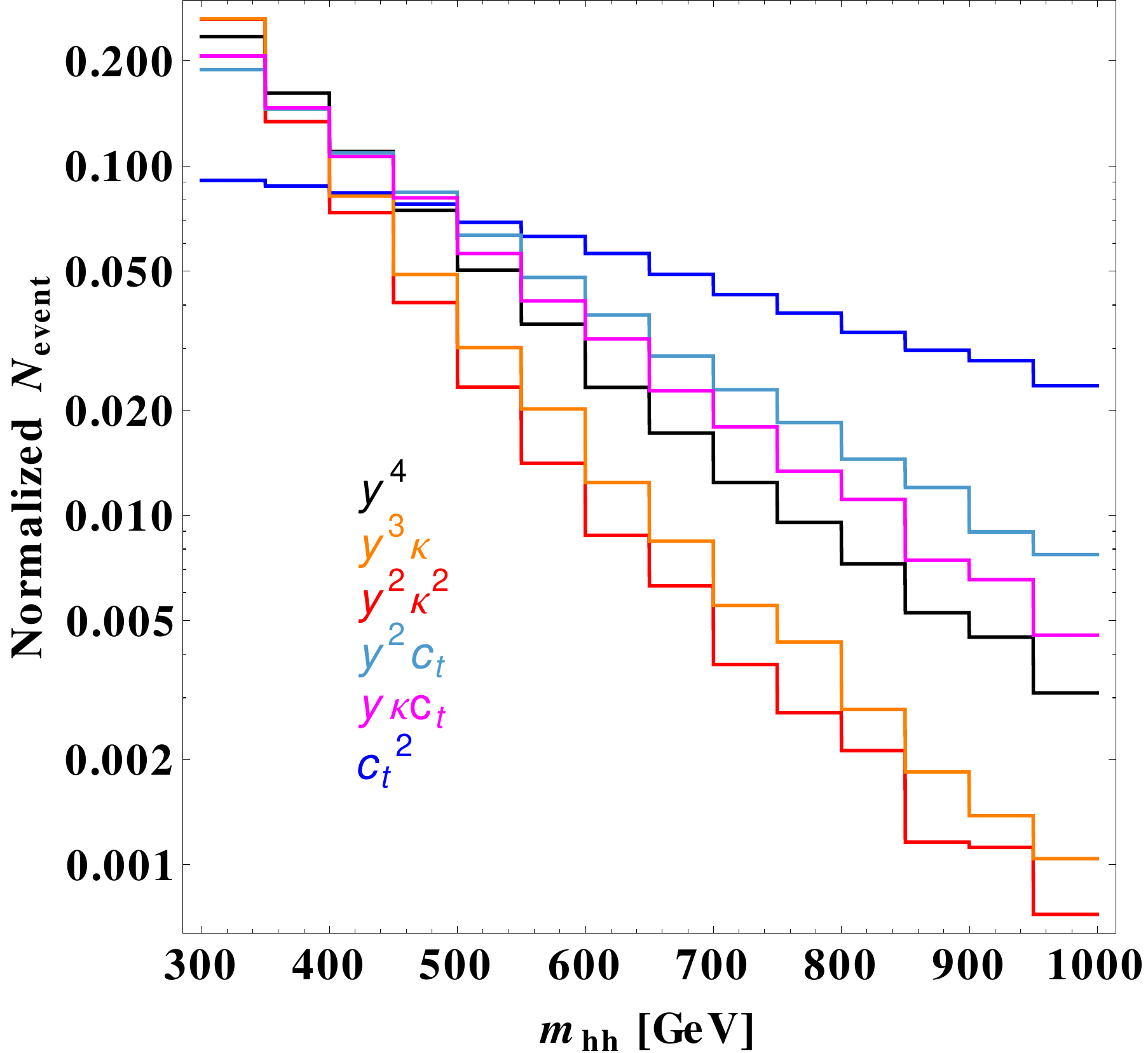}
\includegraphics[scale=0.2]{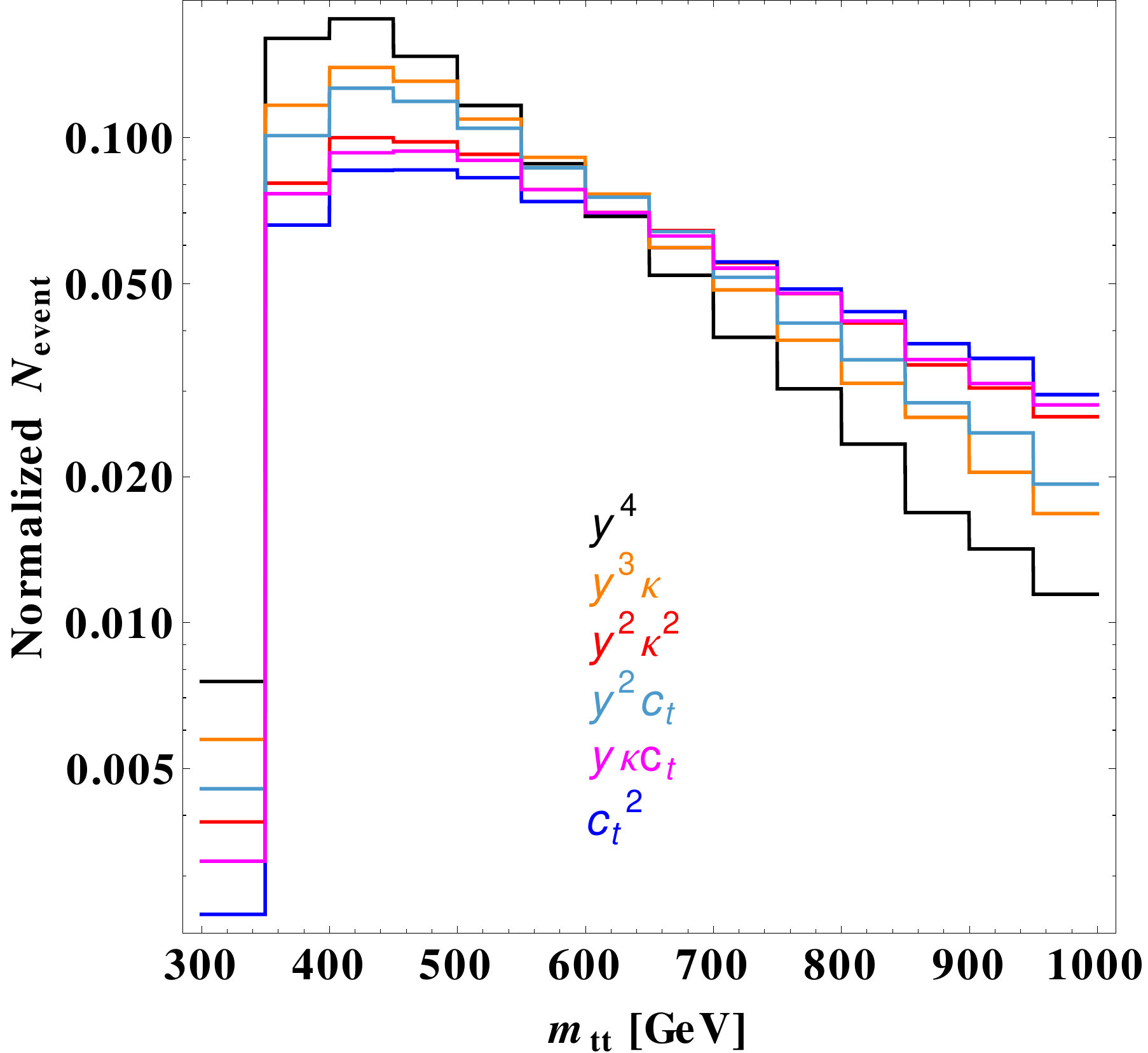}
\includegraphics[scale=0.2]{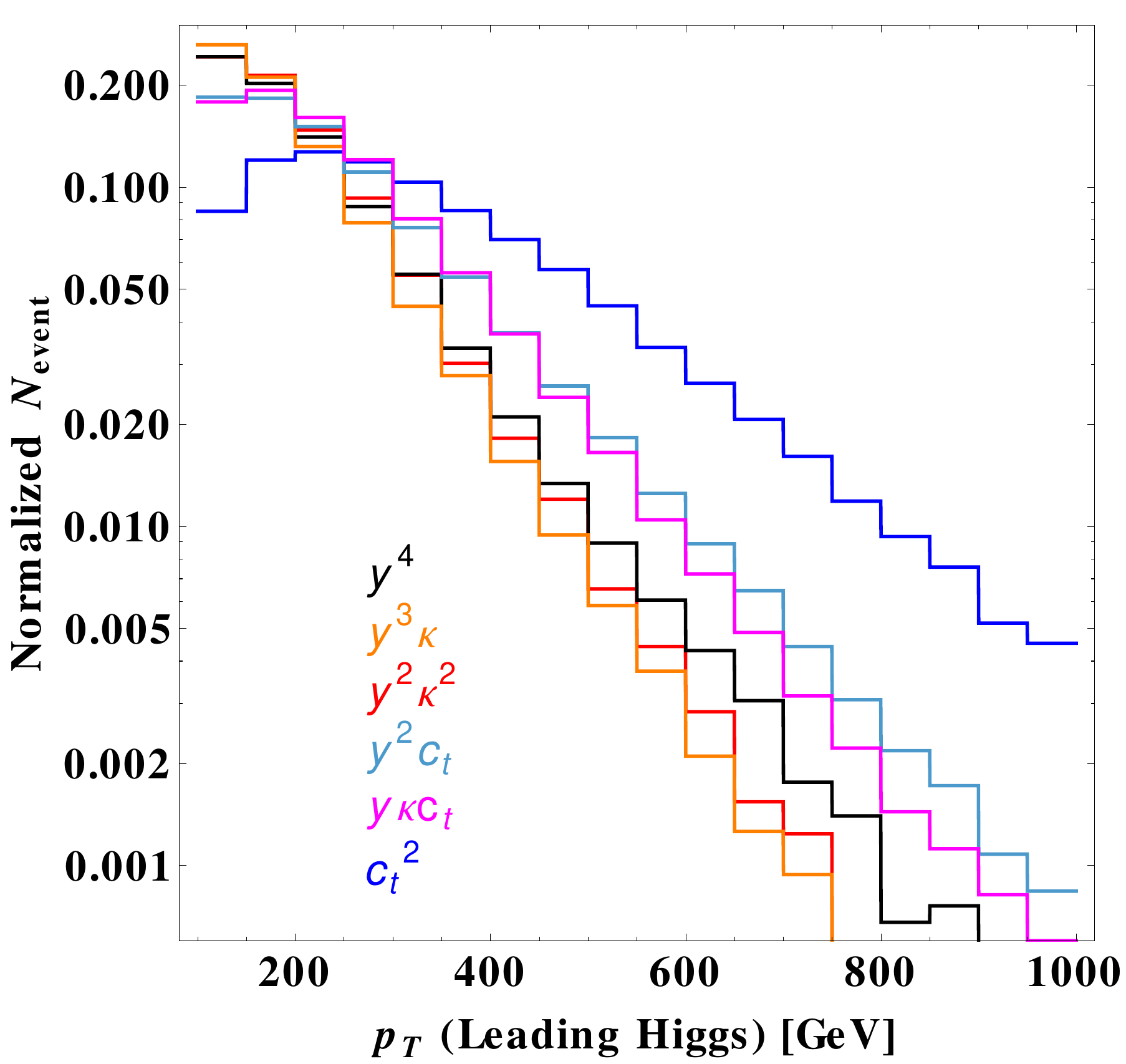}
\includegraphics[scale=0.2]{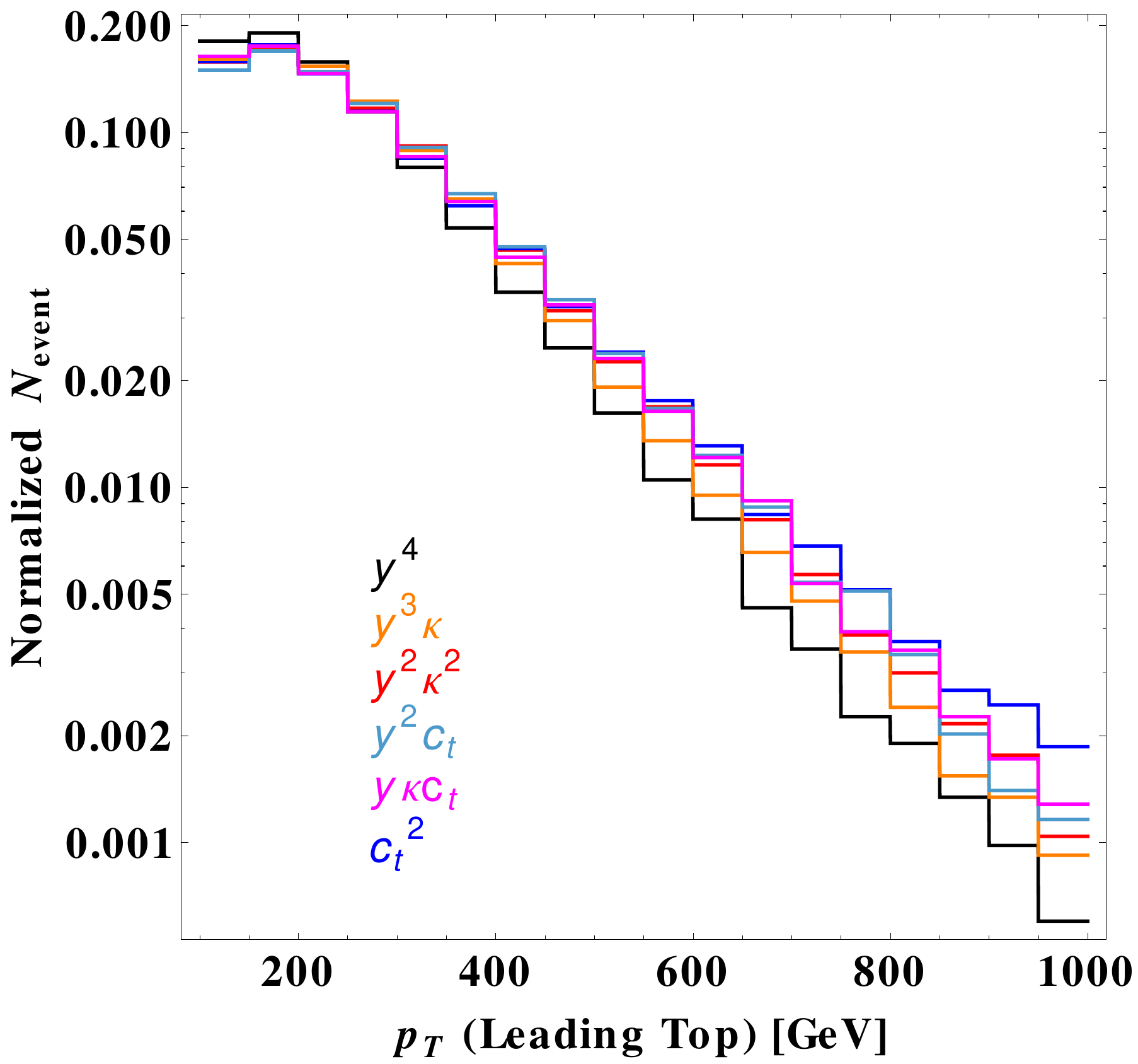}
\includegraphics[scale=0.2]{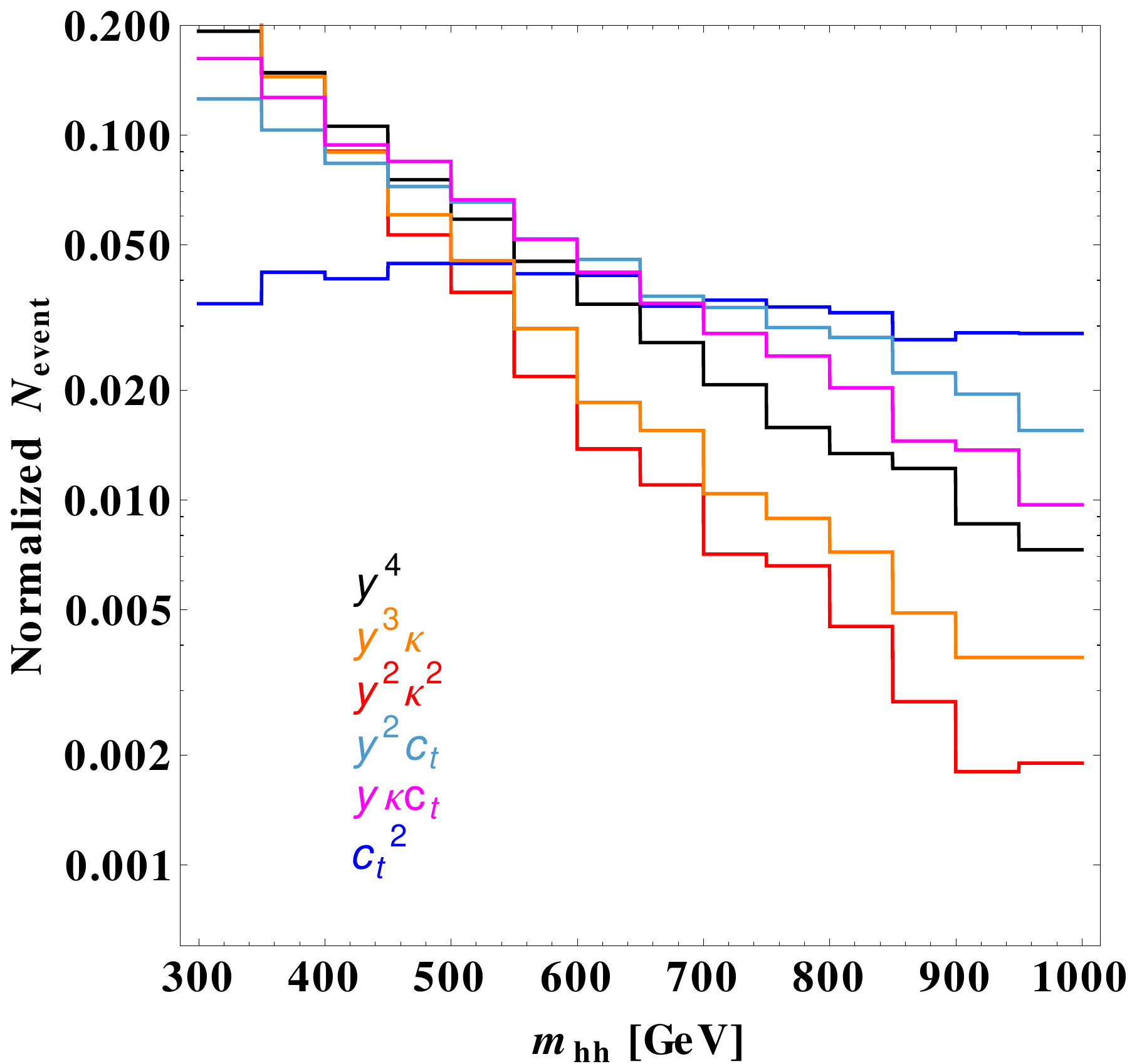}
\includegraphics[scale=0.2]{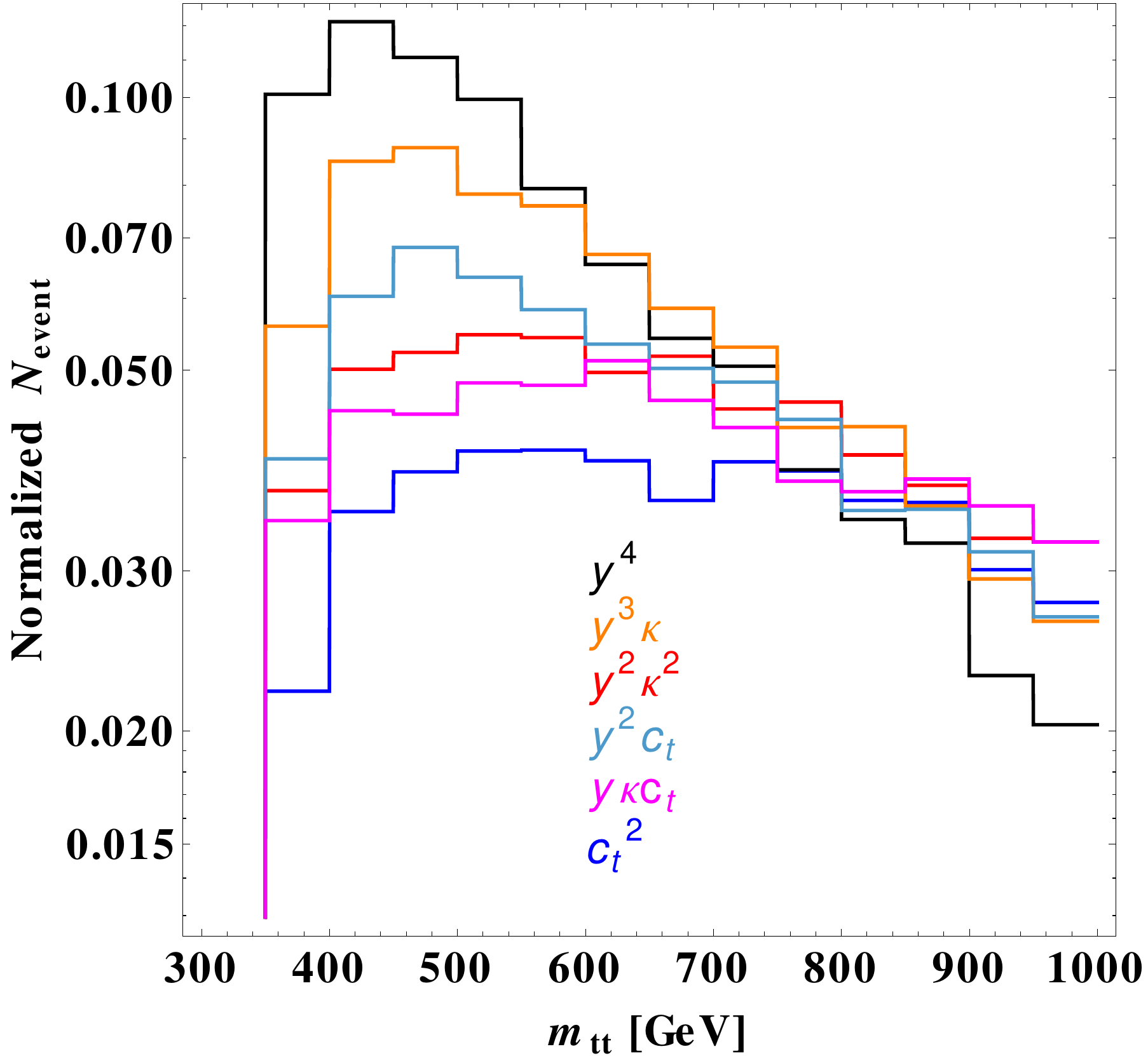}
\includegraphics[scale=0.2]{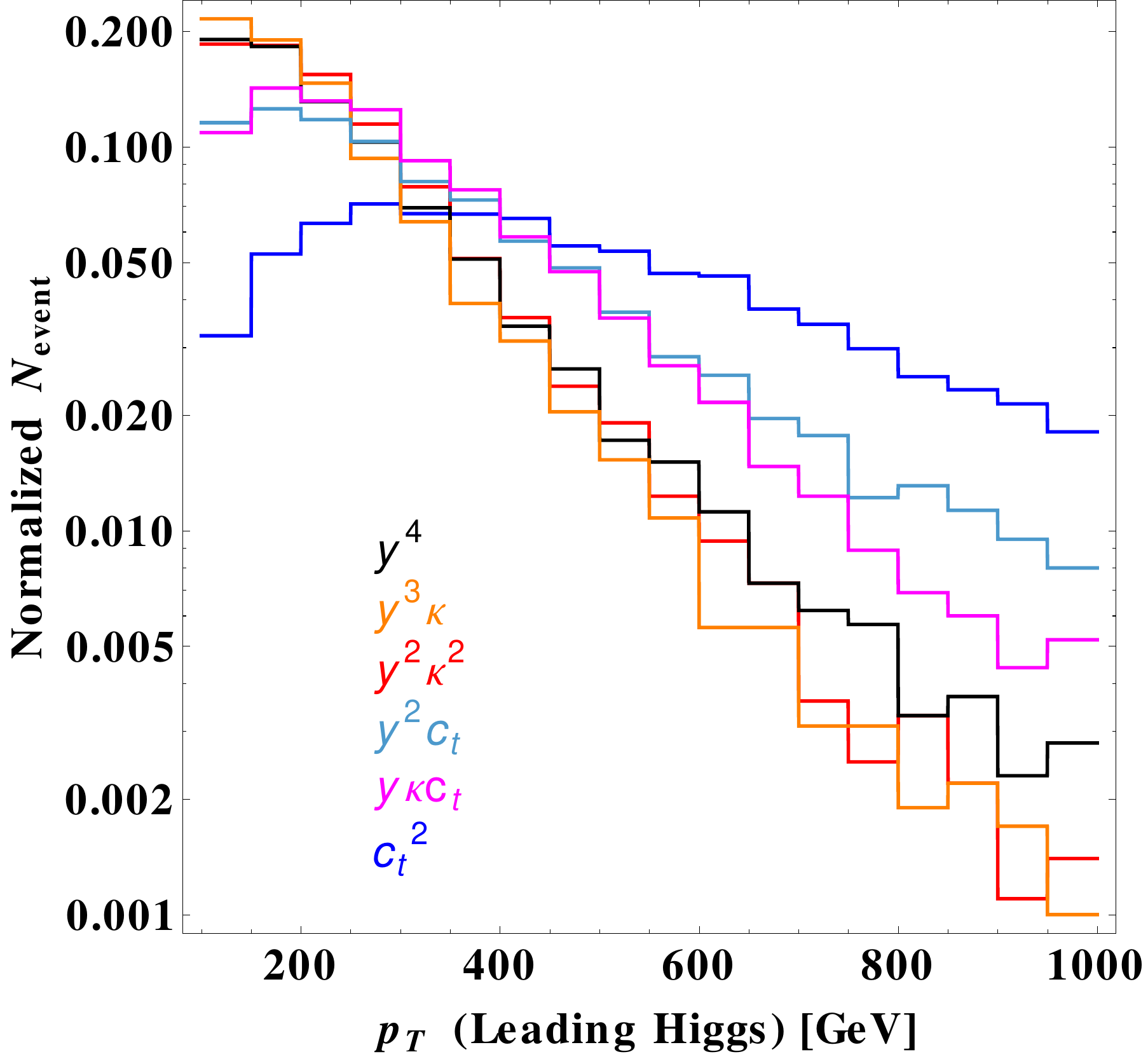}
\includegraphics[scale=0.2]{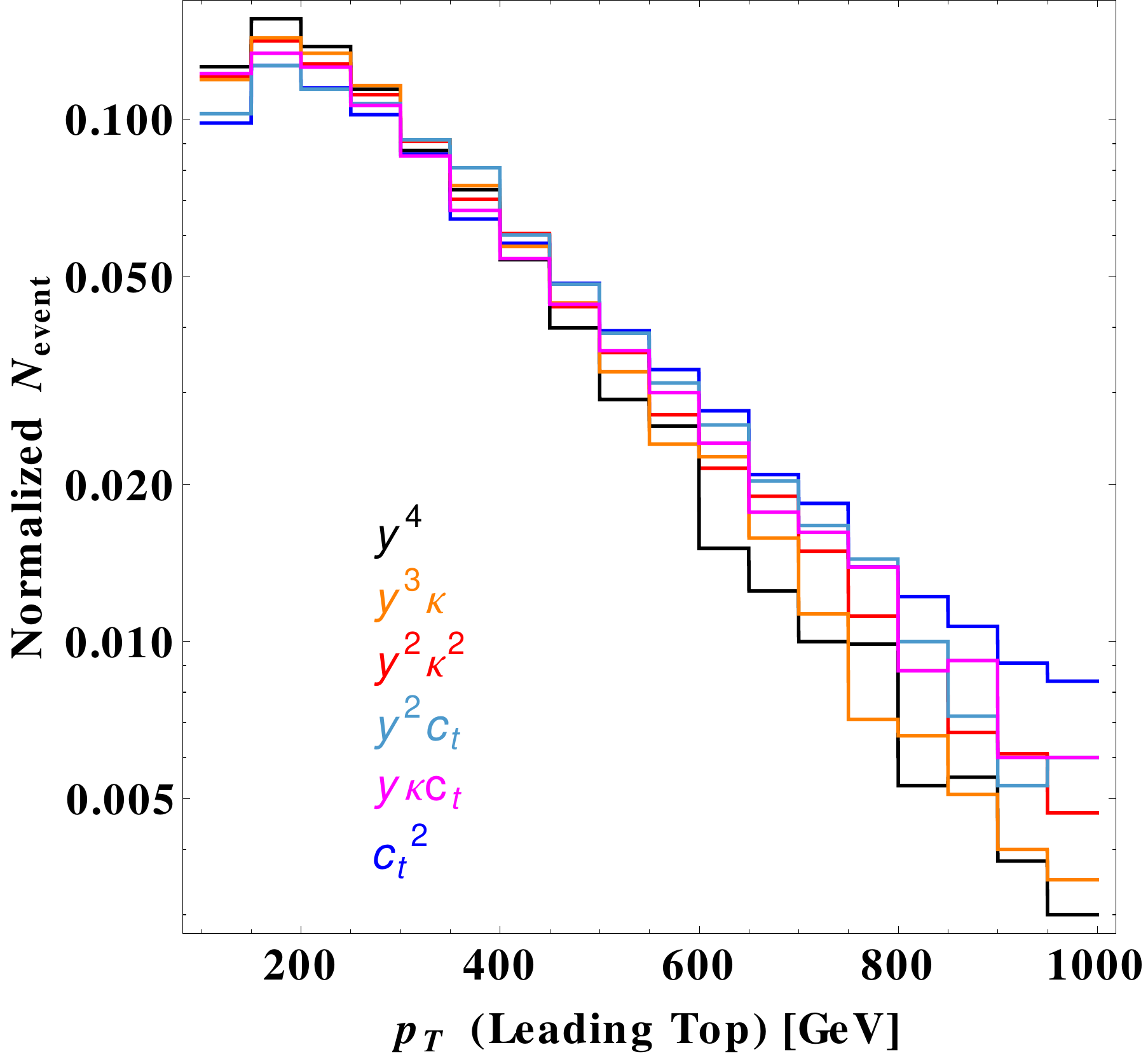}
\caption{Normalized kinematic distributions of the $\{y^4, y^2\kappa^2, c_{t}^2, y^3\kappa, y^2c_{t}, y\kappa c_{t}\}$ samples at parton level, at 14 TeV (top) and 100 TeV (bottom).}
\label{fig:kin2}
\end{figure}

Compared to 14 TeV, the $tthh$ analyses may benefit more from the $y^2\kappa^2$- and $c_t^2$-induced kinematics at 27 TeV and 100 TeV. To make this clear, we define the weights of the $y^2\kappa^2$ term (denoted as $\sigma_{y^2\kappa^2}$) and the $c_t^2$ term (denoted as $\sigma_{c_t^2}$) in $\sigma(tthh)$ as  
\begin{eqnarray}
W_{y^2\kappa^2}(\kappa,c_t) = \frac{\sigma_{y^2\kappa^2}}{\sigma(tthh)}, \ \ \ \ W_{c_t^2}(\kappa,c_t) = \frac{\sigma_{c_t^2}}{\sigma(tthh)} \ , 
\end{eqnarray}
and show their performance at $\sqrt{s}=14$ TeV, 27 TeV and 100 TeV in Figure~\ref{fig:kin1}, with Eq.~(\ref{eq:sig14}-\ref{eq:sig27100}). It is easy to see from this figure that, deviating from $\kappa = 0$ or $c_t = 0$, the weights $W_{y^2\kappa^2}(\kappa, 0)$ and $W_{c_t^2}(1,c_t)$ quickly increase to a level of  $\mathcal O(0.01)$ or $\mathcal O(0.1)$. This could be further improved by event selection. As $\sqrt{s}$ increases, these two weights are gradually enhanced except the region near  $\kappa = 0$ or $c_t = 0$. As a matter of fact,  we have
\begin{eqnarray}
W_{y^2\kappa^2}(1, 0)_{14}  = 0.04, \ \  W_{y^2\kappa^2}(1, 0)_{27}  = 0.05, \ \ W_{y^2\kappa^2}(1, 0)_{100}  = 0.09 
\end{eqnarray}
in the SM limit. As for the weight $W_{c_t^2} (\kappa,c_t)$, its value is suppressed in the SM limit. Yet, with a small deviation in $c_t$, e.g., $c_t=0.5$, we have   
\begin{eqnarray}
W_{c_t^2}(1, 0.5)_{14}  = 0.08, \ \  W_{c_t^2}(1, 0.5)_{27}  = 0.14, \ \ W_{c_t^2}(1, 0.5)_{100}  = 0.25 \ .
\end{eqnarray}
In contrast, such an enhancement w.r.t. beam energy is vague for $\sigma(gg\to hh\to bb\gamma\gamma)$ in Eq.~(\ref{eq:gg1}) and Eq.~(\ref{eq:gg2}). 

In Figure~\ref{fig:kin2}, we show normalized kinematic distributions of the $\{y^4, y^2\kappa^2, c_{t}^2, y^3\kappa, y^2c_{t}, y\kappa c_{t}\}$ samples at 14 TeV and 100 TeV. 
Compared to the others, the $y^2\kappa^2$ and $c_t^2$ samples tend to be separated more from the $y^4$ one in kinematics. They are away from the $y^4$ sample in opposite directions w.r.t. $m_{hh}$ and the leading Higgs $p_T$, while in the same direction w.r.t. $m_{tt}$ and the leading top $p_T$. This can be explained as follows: the two Higgs bosons in the $y^2\kappa^2$ event are produced via off-shell Higgs decay, and hence tend to be soft and central; differently, the two Higgs bosons in the $c_t^2$ event are generated more energetically, resulting in a large $m_{hh}$. These separations become even wider from 14 TeV to 100 TeV.  

A combination of the discussions above justifies that we take a 2D BDT strategy for the analyses at HL-LHC which are based on the $y^2\kappa^2$ and $c_t^2$ samples, 
and meanwhile, raises the expectation that the $y^2\kappa^2$- and $c_t^2$-induced kinematics may further improve the signal efficiency and background rejection $\{\epsilon_{\rm sig}^{ij}, \epsilon_{\rm bg}^{i}\}$ at 27 TeV, and even more at 100 TeV. 

As for the backgrounds of the $tthh$ analyses, the $4t$ cross section increases faster as the beam energy raises (by a factor of $\sim 8.6$ (252) for 27 (100)~TeV) than most of the $tt+X$ ones (by a factor of $\sim 5.7$ (105) for 27 (100)~TeV), because of its relatively higher energy threshold of production. Then, we are able to calculate the projected exclusion limits, assuming that the signal efficiency and background rejection $\{\epsilon_{\rm sig}^{ij}, \epsilon_{\rm bg}^{i}\}$ obtained from the 14 TeV simulations are not changed at 27 TeV and 100 TeV.  
 
\begin{figure}[th]
\centering
\includegraphics[scale=0.35]{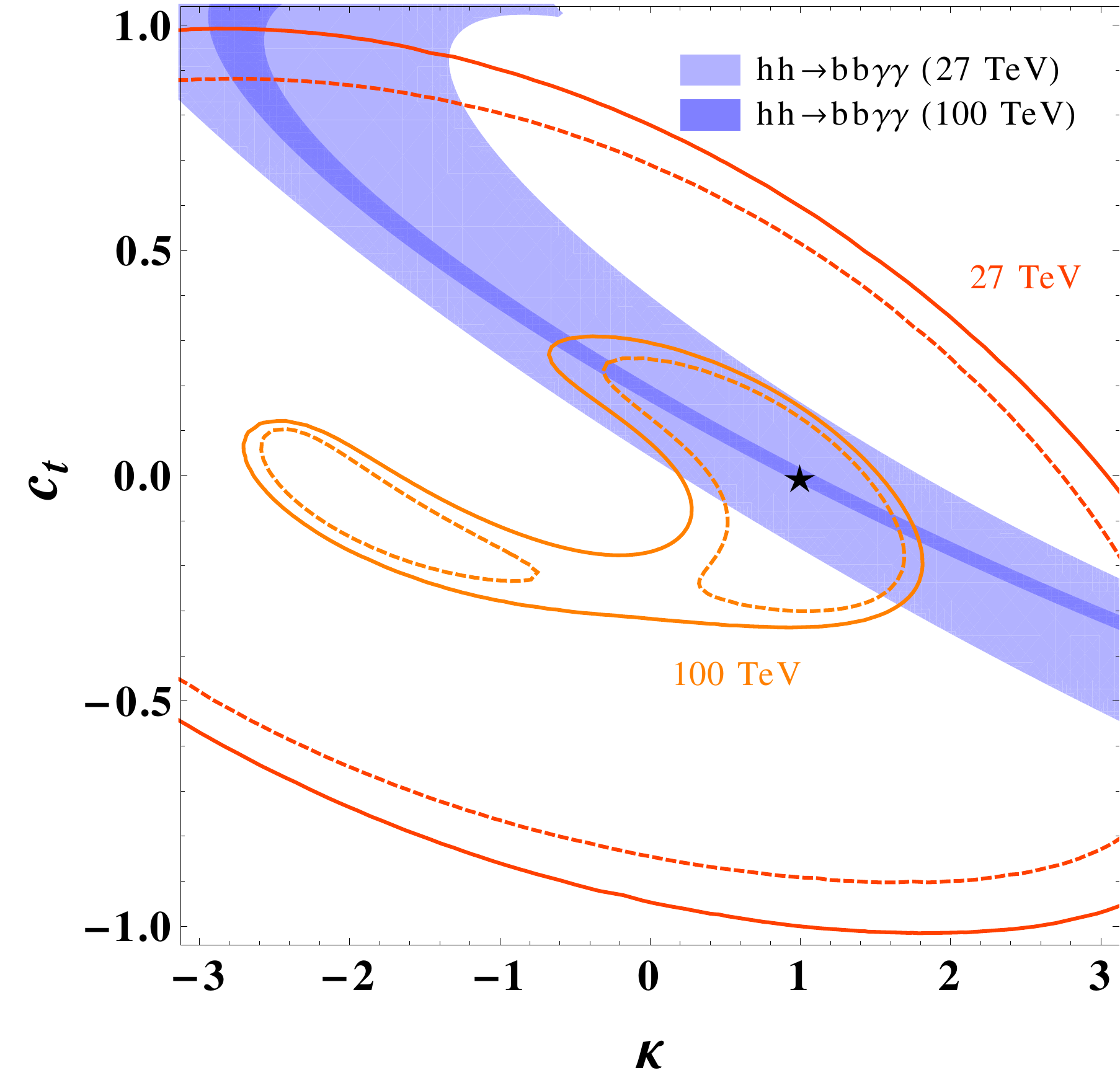}
\caption{Projected exclusion contours in terms of $\kappa$ and $c_t$ at 27 TeV (15 ab$^{-1}$) and 100 TeV (30 ab$^{-1}$), against the hypothesis of background (including $tthh$(SM)) + signal (deviation from $\{\kappa, c_t\} = \{1,0\}$). The solid (dashed) contours are based on a tight (softened tight) $b$-tagging efficiency of 60\% (70\%). 
}
\label{fig:future}
\end{figure}
The projected sensitivities are presented in Figure~\ref{fig:future}. In this Figure, the light blue and blue belt regions represent the sensitivity reaches of the reference analysis of $gg\to hh \to bb \gamma\gamma$ at 27 TeV and 100 TeV, respectively. We notice the difference of the 27 TeV sensitivity in literatures (a precision of $\sim 30\%$ in~\cite{Goncalves:2018yva} and $\sim 80\%$ or worse in~\cite{Homiller:2018dgu,ATL-PHYS-PUB-2018-053} in measuring $\kappa$ at $2\sigma$ C.L.), and a comparison made between them in~\cite{Cepeda:2019klc}. Given that the pileup effect and more backgrounds were simulated in~\cite{ATL-PHYS-PUB-2018-053}, when projecting the 27 TeV sensitivity to the $\kappa - c_t$ plane using Eq. (\ref{eq:gg2}), we require the exclusion contours to reduce in the limit of $c_t=0$ to the results in~\cite{ATL-PHYS-PUB-2018-053}. As for the 100 TeV contours, they are required to reduce to the results in~\cite{Goncalves:2018yva}. The sensitivities generated for the reference analysis by this method are not optimized with kinematics. With the assistance of kinematics, the belt regions are expected to shrink to circular ones~\cite{Azatov:2015oxa}. Despite of this, such a treatment is sufficient for the reference purpose in this analysis. In this figure, the 27 TeV and 100 TeV exclusion contours of $tthh$ are presented in red and orange, respectively. Compared to the 14 TeV (see Figure~\ref{fig:exconts}) and 27 TeV contours, both of which are elliptical, the 100 TeV ones are deformed from left-upper to right-bottom. This deformation is caused by the offset of the SM point from the $\sigma(tthh)$ contour center which is defined by $\{\kappa, c_t\} = \{0,0\}$, together with that the exclusion contours are defined against the SM prediction. At 14 TeV and 27 TeV, the collider sensitivities are relatively low. The scales of their exclusion contours are much larger than this offset. Thus its impact for the contour profile is negligibly small. In the limit of $c_t =0$, the projected $tthh$ sensitivities are worse than that of the reference analysis by a factor of $2-3$. Considering that these sensitivities have not been optimized by, e.g., utilizing the $y^2\kappa^2$- and $c_t^2$-induced kinematics, we would view the projected sensitivities to be encouraging.  

\section{Resonant $tthh$ Analysis}
\label{sec:resultresonant}

\subsection{Heavy Higgs Boson: $ttH \to tthh$}

In type II THDM, physics at the decoupling limit~\cite{Craig:2013hca,Bernon:2015qea,Djouadi:2015jea} provides a natural organizing principle for the collider searches of the heavy Higgs boson $H$. Because its couplings with $WW$, $ZZ$ and $hh$ are highly suppressed, the $H$ is produced and decays mainly through its Yukawa vertices $bbH$ and $ttH$, at high and low $\tan\beta$ regions, respectively. It has been shown that its exclusion limits could be pushed up to $\sim 1$ TeV at HL-LHC and nearly one order more at 100 TeV for moderate and low $\tan\beta$ regions~\cite{Hajer:2015gka, Craig:2016ygr}, using the decay mode of $H\to tt$~\footnote{Precisely speaking, in~\cite{Hajer:2015gka, Craig:2016ygr} the contributions of both CP-even and CP-odd Higgs bosons ($i.e.$, $H$ and $A$) were taken into account in the sensitivity analysis, with an assumption of $m_H=m_A$.} and a BDT method. However, as its mass decreases to $2m_h < m_H < 2 m_t$, the $H$ becomes nearly decoupled. Its decay mode of $H \to tt$ is kinematically forbidden, whereas its couplings with $WW$, $ZZ$ and $hh$ may not be negligibly small. Especially, in the low $\tan\beta$ region the $Hhh$ coupling is larger than the $HWW$ and $HZZ$ ones by an approximate factor $m_H^2/m_Z^2$~\cite{Djouadi:2015jea}, which could result in a branching ratio of $H\rightarrow hh$ larger than 50\%~\cite{Djouadi:2015jea}. A natural expectation is then: the $gg\to H\to hh$ production will generate a sensitivity to this region. Yet, because of the low-$\tan\beta$ enhancement for the $ttH$ coupling, the $H$ could be produced efficiently in association with a pair of top quarks also. The $ttH \to tthh$ analysis thus may provide a probe for this parameter region, alternative to $gg\to H\to hh$.

\begin{table}
\centering
\resizebox{\textwidth}{!}{
\begin{tabular}{|c|c|c|c|c|c|c|}
\hline
 & 5$b$1$\ell$ (fb) & 5$b$2$\ell$ (fb) & SS2$\ell$ (fb) & Multi-$\ell$ (fb) & $\tau\tau$ (fb) & Combined (fb) \\ 
\hline \hline 
$ttH,~m_H=300$~GeV & 3.6 (2.4) & 10 (7.4) & 6.8 (6.5) & 9.2 (8.9) & 12 (11) & 2.5 (2.2) \\ 
\hline 
$ttH,~m_H=500$~GeV &  2.6 (2.0)  & 7.6 (5.7)  & 5.3 (5.1) & 7.4 (7.2)  & 8.0 (7.7) &  2.0 (1.6)  \\ 
\hline 
\hline
$TT,~m_T=1500$~GeV & 0.33 (0.27) & 1.4 (1.2) & 0.87 (0.81) & 1.1 (1.0) & 1.5 (1.5) & 0.24 (0.21) \\ 
\hline 
$TT,~m_T=1750$~GeV & 0.31 (0.25) & 2.4 (1.5) & 0.64 (0.62) & 0.87 (0.83)  & 1.4 (1.4) & 0.20 (0.17) \\ 
\hline 
$TT,~m_T= 2000$~GeV & 0.35 (0.28)  &  3.0 (2.0)  & 0.63 (0.59)  &  1.0 (0.94) & 2.0(1.8) & 0.22 (0.19) \\ 
\hline
\end{tabular} 
}
\caption{Exclusion limits of the $ttH \to tthh$ and $TT \to tthh$ analyses at HL-LHC. The numbers outside (inside) the brackets are based on a tight (softened tight) $b$-tagging efficiency of 60\% (70\%).}
\label{tab:htp}
\end{table}

Note, in general type II THDM the tree-level $HWW/HVV$ couplings depend on the mixing angle of the Higgs sector which is a free parameter. For concreteness, we will fix its value using its tree-level relation with $m_H$ and $\tan\beta$ in the analyses, as is in the MSSM. We will use MadGraph 5~\cite{Alwall:2011uj} to calculate the $H$ production cross section and HDECAY~\cite{Djouadi:1997yw} to calculate its decay branching ratios. Two benchmark points ($m_H=300$, $500$ GeV) are simulated with a BDT strategy described in subsection~\ref{ssec:BDT}. The BDT model is trained against the SM backgrounds including the $tthh$(SM) events at each point (see Table~\ref{tab:BDTSamples}). The model-independent exclusion limits in the five exclusive analyses at HL-LHC and their combination are presented in Table~\ref{tab:htp}. The same as before, the best sensitivities result from the $5b1\ell$ and SS$2\ell$ analyses. These analyses may further gain from using a softened tight $b$-tagging efficiency. A combination of them results in an exclusion limit of $\sim \mathcal O(1)$ fb for the $ttH\rightarrow tthh$ production with $300 \  {\rm GeV} < m_H < 500 \ {\rm GeV}$.

\begin{figure}[th]
\includegraphics[scale=0.5]{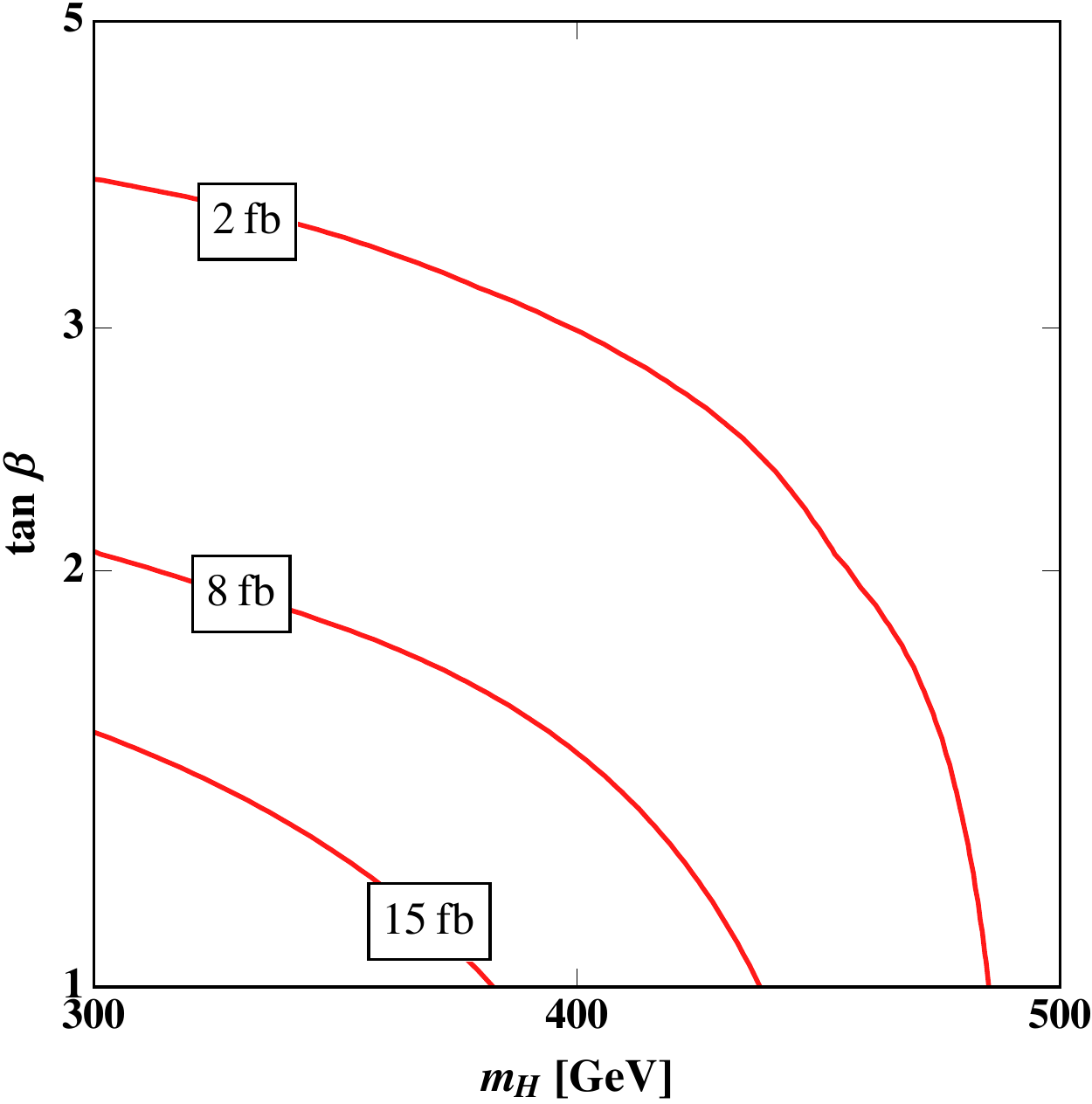} \ \ \ \ 
\includegraphics[scale=0.5]{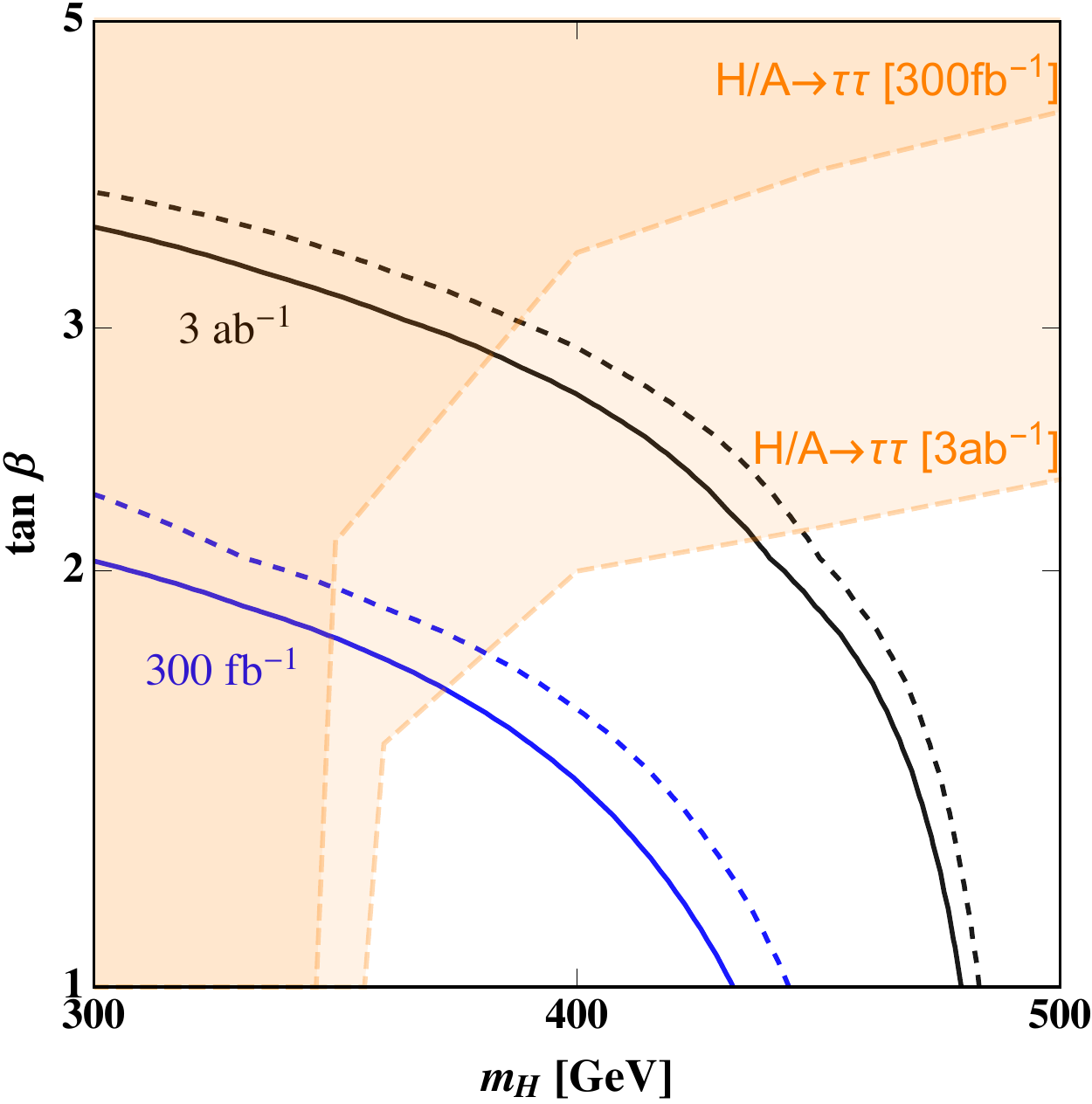}
\centering
\caption{$\sigma(ttH\to tthh)$ contours (left) and combined exclusion limits of the $ttH \to tthh$ analyses (right), with a luminosity of 300 fb$^{-1}$ (blue) and 3 ab$^{-1}$ (black). The solid (dashed) blue and black contours are based on a tight (softened tight) $b$-tagging efficiency of 60\% (70\%). The exclusion limits of the $H/A \to \tau^+\tau^-$ analysis are based on~\cite{Djouadi:2015jea}.}
\label{fig:hh}
\end{figure}

The interpretation of the combined sensitivities in Type II THDM is shown in Fig.~\ref{fig:hh}. As is expected, the $tthh$ analyses can efficiently probe for the nearly decoupling limit with $2m_h < m_H < 2 m_t$ and relatively low $\tan\beta$. For $\tan\beta \sim 1$, the exclusion limit with a luminosity of 300 fb$^{-1}$ goes beyond the $tt$ threshold, reaching $m_H \sim 430$~GeV, with a tight $b$-tagging efficiency of 60\%. The limit can be pushed up to $m_H \sim 480$ GeV at HL-LHC, fully complementing the $bbH/A \to bb \tau\tau$ measurement for the relevant $m_H$ range. Encouragingly, such a sensitivity seems not bad, compared to the $gg\to H \to hh$ one~\cite{Djouadi:2015jea} (also see~\cite{Adhikary:2018ise}), though the latter gains a lot from its relatively large cross section. 

\subsection{Fermionic Top Partner: $TT \to tthh$}
Vector-like top partners extensively exist in the CHM. Because they can be produced via QCD interactions at LHC, the exclusion limits for these particles, either a $SU(2)$ singlet or doublet, have been pushed to above $1.3$~TeV~\cite{Aaboud:2018pii}. As is shown in~\cite{Aaboud:2018xuw}, one of the the major channels involved in this effort is $TT \to tthh$. Below we will analyze its sensitivity at HL-LHC.

\begin{figure}[th]
\centering
\includegraphics[scale=0.4]{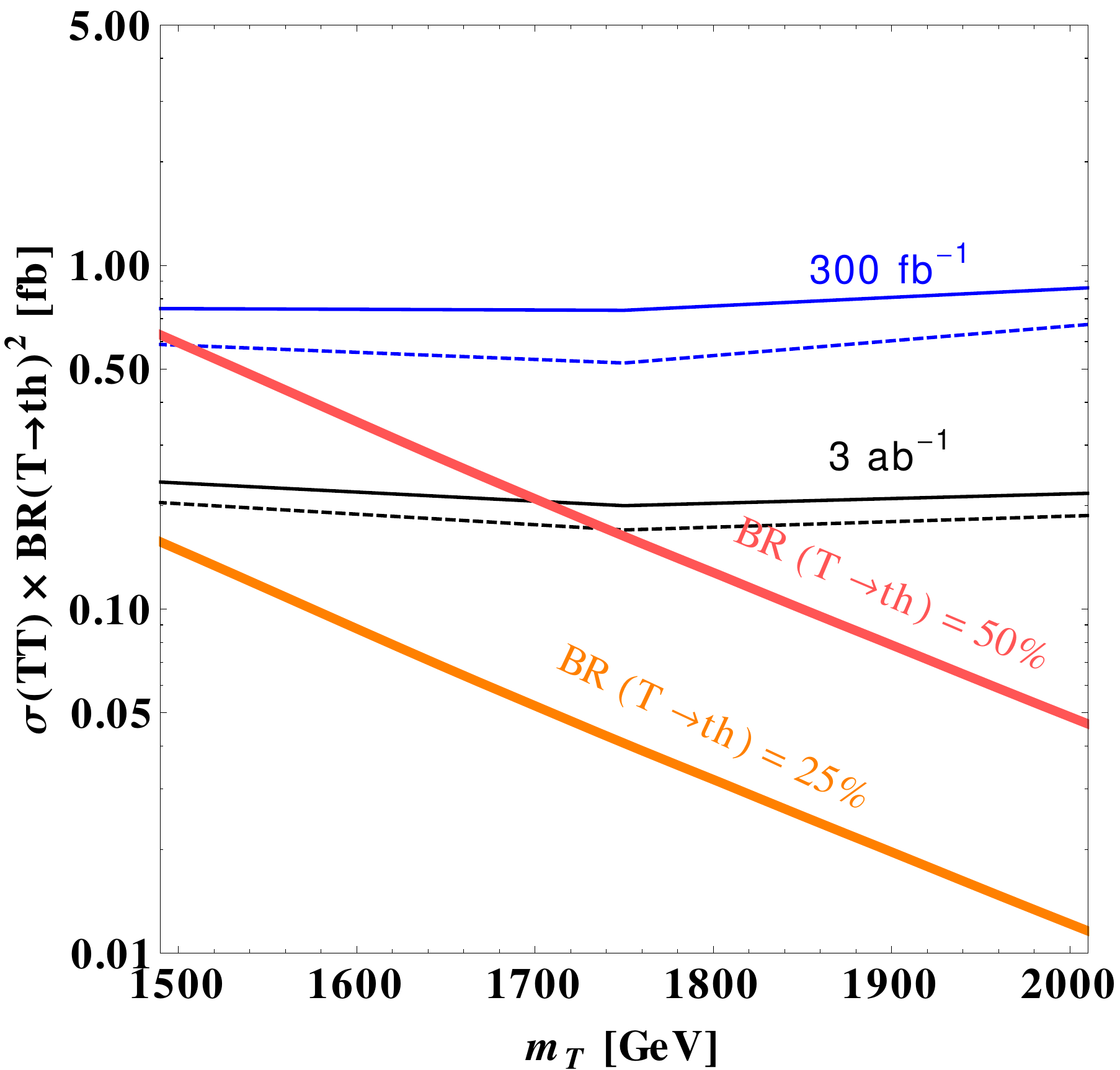}
\caption{Combined exclusion limits of the $TT \to tthh$ analyses, with a luminosity of 300 fb$^{-1}$ and 3 ab$^{-1}$. The solid (dashed) blue and black contours are based on a tight (softened tight) $b$-tagging efficiency of 60\% (70\%). The yellow and red curves are theoretical predictions in two benchmark scenarios which are defined by different branching ratios of $T\to th$.}
\label{fig:tp}
\end{figure}

We simulate three benchmark points with $m_T=1.5$, $1.75$ and 2.0 TeV, with a BDT strategy described in subsection~\ref{ssec:BDT}. As is indicated in Table~\ref{tab:BDTSamples}, the BDT model is trained at $m_T=1.5$ and 2.0 TeV. The analyses at $m_T=1.75$ TeV are done using the BDT model trained at $m_T =1.5$ TeV.
The model-independent exclusion limits in the five exclusive analyses at HL-LHC and their combination are presented in Table~\ref{tab:htp}. As the signal events are characterized by a large $H_T$ (largely due to its large $m_T$), $i.e.$, the scalar sum of the jet $p_T$, they can be well-separated from the backgrounds. Thus, the $TT$ cross section is strongly  constrained to be smaller than $0.20-0.25$ fb, with a weak dependence on $m_T$. The combined exclusion limits are also shown in Figure~\ref{fig:tp}, with a luminosity of 300 fb$^{-1}$ and 3 ab$^{-1}$. Two theoretical benchmark scenarios are projected to this figure. They are characterized by different Br$(T\to th)$, $i.e.$, 25\% and 50\%, which could be achieved for a singlet and doublet top partner, respectively (see, e.g.,~\cite{AguilarSaavedra:2009es}). At HL-LHC, the top partners in these two scenarios could be excluded up to $m_T \sim 1.5$ TeV and $m_T \sim 1.7$ TeV, respectively. Since no exclusive $TT \to tthh$ analysis at LHC is known to us, for comparison we scale the exclusion limit at $m_T=1.43$ TeV, obtained in the $TT \to th +X$ analysis at ATLAS, with a luminosity of $36.1$ fb$^{-1}$ at $\sqrt{s}=13$ TeV~\cite{Aaboud:2018xuw} and an assumption of Br$(T\rightarrow th)=100\%$, to 3 ab$^{-1}$ using Gaussian statistics. This yields a projected limit of $\sigma(TT) < 0.33$ fb, higher than the one at $m_T =1.5$ TeV by a factor $\sim 1.4$.

\section{Conclusion}
\label{sec:conclusion}

In this article, we systematically studied the $tthh$ physics at HL-LHC, using the five exclusive analyses: $5b1\ell$, $5b2\ell$, SS2$\ell$, multi-$\ell$ and $\tau \tau$ resonance. To address the complication of event topologies and the mess of combinatorial backgrounds, the BDT method was applied. We showed that the $tthh$ production can serve as a useful probe for the cubic Higgs self-coupling, the $tthh$ contact interaction, the new resonances such as heavy Higgs boson and fermion top partners. The measurement results may assist a further study on the nature of EWPT and EWSB. The main conclusions include: 
\begin{itemize}
 
\item In addition to $5b1\ell$ which was first suggested in~\cite{Englert:2014uqa,Liu:2014rva}, the SS2$\ell$ (or SS2$\ell+$multi-$\ell$) analysis provides another promising channel to study the $tthh$ physics. For the $tthh$(SM) production at HL-LHC, the BDT sensitivity of $5b1\ell$ is $\sim 30\%$ better than that of SS2$\ell$, and comparable to that of SS2$\ell+$multi-$\ell$, with a tight $b$-tagging efficiency of 60\%. But, its $S/B$ value is $\sim 25\%$ smaller than that of SS2$\ell$ or SS2$\ell+$multi-$\ell$. A  combination of these exclusive analyses potentially allows for detecting $tthh$(SM) with a statistical significance $\sim 0.9\sigma$. The $5b1\ell$ sensitivity could be further improved with a softened tight $b$-tagging efficiency. But, in this case, the backgrounds $tt+ 4c, 2b/2c2j, 4j$ need to be simulated in a more solid way.   

\item The $tthh$ analyses at HL-LHC can be applied to probe for anomalous cubic Higgs self-coupling and the $tthh$ contact interaction. To utilize the kinematic features brought in by both, we defined a 2D BDT framework. Along each of its dimensions one specific signal sample ($i.e.$, the samples of $y^2\kappa^2$ and $c_t^2$) is favored. 
With this method we showed that a combination of the five exclusive analyses potentially can constrain $\kappa$ to $\{6.9, -10\}$ for $c_t=0$, and $c_t$ to $\{-2.7,1.8\}$ for $\kappa = 1$. This may partially address the sensitivity degeneracy w.r.t. the cubic Higgs self-coupling, a difficulty usually thought to exist in the gluon-fusion di-Higgs analysis at HL-LHC.

\item To qualitatively measure the potential of the future hadron colliders, we projected the non-resonant $tthh$ sensitivities at HL-LHC  to 27 TeV and 100 TeV, assuming the signal efficiency and background rejection at HL-LHC to be unchanged. We showed that a combination of the five exclusive analyses could detect the $tthh$(SM) production with a significance of $3.1\sigma$ and $14.3\sigma$ at 27 TeV and 100 TeV, respectively, and constrain $\kappa$ to $\{-2.6, -1.6\} \cup \{ 0.2, 1.6\}$ for $c_t=0$, and $c_t$ to $\{-0.34, 0.16\}$ for $\kappa = 1$, at $2\sigma$ C.L. at 100 TeV~\footnote{In a recent paper~\cite{Banerjee:2019jys}, the authors analyzed the $5b1\ell$ sensitivity at 100 TeV. Compared to it, the projected sensitivities obtained here are roughly better by a factor of two in both $\kappa$ and $c_t$ directions. Especially, we showed that the $tthh$ analyses have a potential to exclude the parameter region around $\{\kappa, c_t \}=\{0,0\}$ in Figure~\ref{fig:sigma27+100}. This could partially benefit from the combination of multiple channels in our analysis.}.

\item The $tthh$ analyses can be also applied to search for new heavy resonances. For illustration, we considered two representative scenarios: $ttH \to tthh$ in type II THDM and $TT \to tthh$ in the CHM. At HL-LHC, the exclusion limit for the $H$ can be pushed beyond the $H\to tt$ threshold, up to $\sim 480$ GeV for $\tan \beta \sim 1$. This complements well the $bbH \to bb\tau\tau$ analysis in the relevant $m_H$ range. In the latter case, the $tthh$ analyses set up an upper limit $\sim 0.2$ fb for the $TT$ cross section with $m_T$ between $1.5 - 2.0$ TeV. This constraint excludes a scenario with Br$(T\to th) = 25\%$ and $50\%$ up to $\sim 1.5$ TeV and $\sim 1.7$ TeV, respectively.    
 
\end{itemize}

We would bring it to attention that the following effects were not systematically taken into account in this study. They may weaken the analysis sensitivities obtained in this article. First, though we introduced a $K$ factor to encode the NLO uncertainty in the background cross sections, we did not explore the impact of systematic errors in detail. Also, we did not simulate the pileup effect directly, though we applied a trimming to fat jets and smeared the MET. Additionally, for a softened tight $b$-tagging efficiency of 70\%, we did not simulate the impact of the $tt+ 4c, 2b/2c2j, 4j$ backgrounds. Though this choice may improve the analysis sensitivities, especially for $5b1\ell$ and $5b 2\ell$, a solid simulation of these backgrounds is necessary.  

The study on the $tthh$ physics will greatly benefit from future hadron colliders. Because of its relatively large energy threshold of prodction, the $tthh$ cross section increases faster as the beam energy increases, compared to some main di-Higgs production mechanisms such as gluon fusion~\cite{Frederix:2014hta}. The enhanced boost in kinematics with a higher beam energy will benefit to reconstructing the signal events also. Furthermore, as $\sqrt{s}$ increases, the weights of the $y^2\kappa^2$ and $c_t^2$ terms in the non-resonant $\sigma(tthh)$ will be gradually enhanced (except some narrow regions), resulting in more contributions from them to the $tthh$ production at 27 TeV and 100 TeV, compared to 14 TeV. This may strengthen the $tthh$ sensitivities at future hadron colliders to the anomalous cubic Higgs coupling and the $tthh$ contact interaction. Together with a potentially larger luminosity, these factors may significantly improve the sensitivities of these $tthh$ analyses. The good sensitivities of these analyses may extend to some new channels, e.g., the one with  di-photon Higgs decay. As a matter of fact, $\sim 10^{4}$ di-photon $tthh$(SM) events, three orders more than those at HL-LHC, can be generated with a luminosity of 30 ab$^{-1}$ at 100 TeV. Given that di-photon Higgs boson can be reconstructed with a high efficiency and its backgrounds are relatively clean, this channel may also play a non-trivial role in studying the $tthh$ physics. Naturally, pursuing a comprehensive study regarding these lays out an ongoing project~\cite{LLLL}. 

The study on the $tthh$ physics, together with the other Higgs measurements, at future colliders may allow a deeper exploration on the nature of Higgs boson, the drive of EWPT, and the underlying theory of EWSB. One example is the reconstruction of Higgs potential. It is well-known that di-Higgs productions can be applied to probe for the cubic and quartic Higgs self-couplings. Yet, the renormalization of the quartic Higgs self-coupling to the cubic one result in a dependence on renormalization scheme, due to incomplete treatment of higher-order effects. As was discussed in~\cite{Liu:2018peg}, a combined analysis of multiple di-Higgs productions can greatly reduce such a scheme dependence at linear level. The $tthh$ production at hadron colliders may well-pair with the $gg\to hh$ production to achieve this. Another example is pinning down the underlying theory of the Higgs self-couplings or the $tthh$ contact interaction, given an anomaly. In either case, such an anomaly could arise from multiple new physics scenarios:  an anomalous cubic Higgs coupling can be induced in the SM extension with, e.g., Higgs singlet, doublet or triplet, while a nonzero $tthh$ contact interaction can arise from the CHM or supersymmetry at either tree or loop level. To recognize the real one, a combination of multiple Higgs probes is necessary.  We leave these explorations to a future work.


\begin{acknowledgments}
We would greatly thank Jing Ren, Zhangqier Wang, Sijun Xu for valuable discussions. This research is supported by the General Research Fund (GRF) under Grant No. 16312716 which was issued by the Research Grants Council of Hong Kong S.A.R.
\end{acknowledgments}

\newpage

\appendix

\section{BDT Variables and Samples}
\label{sec:BDTv}

\begin{table}[th]
\centering
\resizebox{\textwidth}{!}{
\begin{tabular}{c|c|c|c|c|c}
\hline
BDT variables & 5$b$1$\ell$ & 5$b$2$\ell$ & SS2$\ell$ & Multi-$\ell$ & $\tau\tau$\\
\hline
$N_{j}$ & \c & \c & \c & \c & \c\\
$H_T$ & \c & \c  & \c & \c & \c\\
MET & \c & \c  & \c & \c & \c\\
$M_T$ & \c & / & / & / & /\\
Leverage~\cite{Cheng:2016npb} & / & \c & \c & \c & /\\
Max/Avg($\Delta \eta_{bb}$) &\c & \c &\c & \c & \c\\
Min/Max/Avg($m_{bb}$) & \c  & \c &\c & \c &\c\\
Min/Avg($R_{b,\ell}$) &\c & \c &\c & \c &\c\\
Centrality$(b/j)$& \c & \c &\c & \c &\c\\
$p_T(b_i)$ &5,6 & 3-6 & 1-4 & 2-4 & 2-4\\
$p_T(j_i)$ & 4 & / & 1-3 & 2 & 2,3\\
$p_T(\ell_i)$ & 1 & 1,2 & 1,2 & 1-3 & 1\\
$\eta(\ell_i)$ & 1 & 1,2 & 1,2 & 1-3 & 1\\
$p_T({\rm Fat}~j_i)$ & 1 & 1 & 1 & 1 & /\\
$m({\rm Fat}~j_i)$ & 1 & 1 & 1 & 1,2 & /\\
\hline
$O(t_i)$ &1,2 &1,2 & 1,2 & 1,2 & 1,2\\
$O(h_i)$ &1-3 &1-3 & 1-3 & 1-3 & 1-3\\
$\Delta R_{t_i,t_j}$ $[m_{t_i,t_j}]$ & / & / & 1,2 & / & /\\
$\Delta R_{t_i,h_j}$ $[m_{t_i,h_j}]$ & i=1, j=1,2 & / & i=1,2, j=1,2 & i=1, j=1,2 & /\\
$\Delta R_{h_i,h_j}$ $[m_{h_i,h_j}]$  & 1-3 & 1-3 & 1-3 & 1,2 & 1,2 \\
$\Delta R_{t_i,\ell_j}$ $[m_{t_i,\ell_j}]$ & i=1, j=1 &i=1, j=1,2 & / & / & /\\
$\Delta R_{h_i,\ell_j}$ $[m_{h_i,\ell_j}]$ & i=1-3, j=1&i=1,2, j=1,2 & i=1,2 j=1,2 & i=1, j=1-3 & i=1,2, j=1\\
$m_{\ell_i\ell_j}$ & / & 1,2 & 1,2 & 1-3 & /\\
\hline
$m_{\tau\tau}$ & /&/ & / & / & \c \\
$p_{T}(\tau\tau/\ell\ell)$ & / & \c & / & / & / \\
$\eta(\tau\tau/\ell\ell)$ & / &\c & / & / & / \\
$\Delta R_{t_i,\tau\tau/\ell\ell}$ $[m_{t_i,\tau\tau/\ell\ell}]$ & / &1,2 & / & / &1\\
$\Delta R_{h_i,\tau\tau/\ell\ell}$ $[m_{h_i,\tau\tau/\ell\ell}]$& / & 1,2 & / & /&1,2\\
$m_{\ell_i,\tau\tau}$ & /  & / & / & / &1 \\
\hline
\end{tabular}
}
\caption{Variables used for the BDT analyses. ``$j$'' represents jets of light flavor. All leptons and jets are sorted by $p_T$. Top quarks and Higgs bosons are sorted by their BDT reconstruction score. In the resonant $tthh$ analyses, the variables of separation for the reconstructed objects (outside the square brackets) are replaced by the variables of invariant mass (inside the square brackets).
}
\label{tab:BDTvariables}
\vspace*{0mm}
\end{table}

\begin{table}[th]
\centering
\resizebox{\textwidth}{!}{
\begin{tabular}{c|c|c|c|c|c}
Training (67\%) + testing (33\%)  & 5$b$1$\ell$ & 5$b$2$\ell$ & SS2$\ell$ & Multi-$\ell$ & $\tau\tau$\\
\hline\hline
Backgrounds (excluding $tthh$(SM)) & 24796 (53298) & 3281 (7005) & 12199 (13149) & 8226 (8778) & 1750 (1896)\\
$tthh$(SM) & 34559 (66065) & 4140 (8140) & 9921 (10659) & 5809 (6263) & 2431 (2602)\\
$y^2\kappa^2$-specific & 9352 (18013) & 1049 (2147) & 2650 (2851) & 1460 (1589) & 628 (674)\\
$c_t^2$-specific  & 9856 (18935) & 1197 (2279) & 2905 (3123) & 1611 (1769) & 826 (906)\\
$ttH,~m_H=300$~GeV & 19934 (36967) & 2137 (4210) & 4173 (4477) & 2320 (2488) & 1130 (1217)\\
$ttH,~m_H=500$~GeV & 11787 (22393) & 1364 (2638) & 2482 (2673) & 1473 (1576) & 822 (881)\\
$TT,~m_T=1500$~GeV & 15634 (30521) & 1573 (3052) & 4001 (4345) & 2259 (2420) & 1249 (1346)\\
$TT,~m_T=2000$~GeV & 10067 (20108) & 791 (1719) & 2721 (2931) & 1374 (1488) & 814 (863)\\
\end{tabular}

}
\caption{Samples used in the BDT analyses. In each sample, $67\%$ of the events are used for training and $33\%$ for testing. The numbers outside (inside) the brackets are based on a tight (softened tight) $b$-tagging efficiency of 60\% (70\%).}  
\label{tab:BDTSamples}
\end{table}

\newpage
\newpage

\bibliography{Biblotthh}

\end{document}